\documentclass[12pt]{article}%
\usepackage{amsmath}
\usepackage{amsfonts}
\usepackage{amssymb}
\usepackage{graphicx,comment}%
\usepackage{booktabs}
\usepackage{setspace}
\usepackage[shortlabels]{enumitem}
\usepackage{comment}

\usepackage[textfont=it]{subcaption}
\usepackage[labelfont=bf]{caption}
\usepackage[utf8]{inputenc}
\usepackage{adjustbox}
\usepackage{natbib}
\usepackage{multirow,array}

\usepackage{float}
\usepackage{ragged2e}

\newcommand{\foldersimu}{inputs}
\newcommand{\foldersimubis}{inputs}
\newcommand{\folderempi}{inputs}

\usepackage[table,xcdraw,dvipsnames]{xcolor}

\usepackage[colorlinks=true]{hyperref}
\hypersetup{
	colorlinks = true,
	linkcolor = Blue,
	anchorcolor = Blue,
	citecolor = Blue,
	filecolor = Blue,
	urlcolor = Blue
}
\usepackage{chngcntr}
\usepackage[titletoc,title]{appendix}
\usepackage[dotinlabels]{titletoc}

\usepackage{pdflscape}

\newcolumntype{P}[1]{>{\centering\arraybackslash}p{#1}}

\usepackage[draft, color = yellow, textsize=tiny]{todonotes}

\definecolor{dkred}{rgb}{0.8, 0.0, 0.0}

\providecommand{\U}[1]{\protect\rule{.1in}{.1in}}

\begin{document}
	
	\title{\Large \textbf{Robust Impulse Responses using External Instruments: the Role of Information}\thanks{This paper has benefited from discussions with Mario Forni and participants at the UEA Time Series Workshop, the Workshop in Empirical and Theoretical Macroeconomics and various research seminars.}} 
        
        
        
        
	
	
	

\author{Davide \textsc{Brignone}\thanks{Università di Roma Tor Vergata, email: davide.brignone@uniroma2.it.}
\and Alessandro \textsc{Franconi}\thanks{Università di Pavia, email: alessandro.franconi@unipv.it	}
\and Marco \textsc{Mazzali}\thanks{Università di Bologna, email: marco.mazzali5@unibo.it.}
}

\bigskip

\date{This version: \today}

\maketitle

\bigskip

\begin{abstract} 

\singlespacing
\noindent External-instrument identification leads to biased responses when the shock is not invertible and the measurement error is present. We propose to use this identification strategy in a structural Dynamic Factor Model, which we call Proxy DFM. In a simulation analysis, we show that the Proxy DFM always successfully retrieves the true impulse responses, while the Proxy SVAR systematically fails to do so when the model is either misspecified, does not include all relevant information, or the measurement error is present. In an application to US monetary policy, the Proxy DFM shows that a tightening shock is unequivocally contractionary, with deteriorations in domestic demand, labor, credit, housing, exchange, and financial markets. This holds true for all raw instruments available in the literature. The variance decomposition analysis highlights the importance of monetary policy shocks in explaining economic fluctuations, albeit at different horizons.\\


\end{abstract}

\newpage

\section{Introduction}
\counterwithout{figure}{section}

The use of external instruments to identify structural shocks has now become a widely used methodology in macroeconomics for estimating dynamic causal effects. The main advantage of employing such identification strategy, compared to traditional schemes, is its independence from any assumption on sign or timing of the impulse responses \citep[see][for a detailed review]{stock2018identification}. 

However, this identification strategy is not immune to problems affecting standard vector autoregressive (VARs) models. As largely discussed by the literature, the reliability of this class of models in estimating the transmission of structural shocks may be undermined by two major limitations: the \textit{curse of dimensionality} and the presence of \textit{measurement errors}. The first is linked to the restricted amount of information available to the econometrician, which is forced to use small- or, at most, medium-scale systems due to the exponentially growing parameter space. Implicitly, this procedure assumes that the information from the past and present observations of the selected endogenous variables is sufficient to recover the structural shock of interest, \textit{i.e.}, the model is \textit{informational sufficient} \citep[following the definition of][]{forni2014sufficient}. However, several studies demonstrated that this assumption does not always hold true, as for instance, in the presence of anticipated shocks, that is, shocks with a delayed effect on some variables \citep[see][among others]{leeper2013fiscal,forni2014no}. Consequently, the issues of \textit{non-invertibility} or \textit{non-fundamentalness} may arise.\footnote{While invertibility requires that the shocks can be inferred from past and current values of the endogenous variables, fundamentalness implies that the shocks only need to belong to the space spanned by those values. Note that while the two properties are closely related, they are not exactly the same thing, although they are often used as synonyms. For example, consider the case where the vector moving average representation of a square system has at least one root equal to one in modulus. The system would not be invertible, but it would still be fundamental. See, \textit{inter alia}, \cite{hansen1991two}, \cite{lippi1993dynamic, lippi1994var} and the review by \cite{alessi2011non}.} A second source of bias may be the presence of a non-negligible idiosyncratic component of the series, which is usually interpreted as measurement error or as sectoral disturbances. Indeed, despite often neglected in the literature, this could further affect the estimates of the impulse response functions, leading to inaccurate results - see, for instance, \cite{giannone2006vars} and \cite{forni2020common}. 

In the last decades, the literature attempted to deal with the curse of dimensionality proposing econometric models that allow to include a larger number of variables \textit{w.r.t.} VARs. A non-exhaustive list of examples includes Bayesian VARs \citep[see, \textit{e.g.},][]{de2008forecasting,banbura2010large}, Reduced-Rank VARs \citep[see,\textit{e.g.},][]{carriero2011forecasting,cubadda2022reduced}, Factor Augmented VARs \citep[FAVAR, henceforth - see, \textit{e.g.},][]{bernanke2005measuring}, and Dynamic Factor Models \cite[DFM, henceforth - see, \textit{e.g.},][]{forni2000generalized,forni2009opening,stock2002forecasting}. Although they manage to solve the missing information problem, only the last approach is able to deal with measurement errors. 

In this paper, we propose using external instruments in a Dynamic Factor Model (Proxy DFM, henceforth). We describe how to apply the external instrument identification in a structural DFM framework to estimate a unit-variance shock. This allows the estimation of the variance decomposition and, possibly, the historical decomposition. By means of the simple theoretical model with perfect foresight studied in \cite{leeper2013fiscal}, we show the benefits of applying the proposed identification strategy within a \textit{data-rich environment} compared to a small-scale (VAR) model. We compare the theoretical responses with those obtained by estimating impulse responses using both a Proxy SVAR and Proxy DFM from the simulated series of the model. We explore various model specifications and highlight the role of measurement error and non-fundamentalness in biasing the estimated results. 

Our findings indicate that, when the information set is insufficient, Proxy VAR delivers biased responses. Conversely, the DFM successfully estimates the true IRFs. Moreover, informational deficiency may be compounded by other sources of bias, such as the sensitivity of estimates to the chosen model specification and the presence of measurement error. For example, even if the VAR is correctly specified and the system is fundamental, the presence of measurement error can bias the final estimates. In contrast, the DFM is not affected by any distortion. Its robustness holds even when higher levels of the idiosyncratic component are added to the simulated series.	

We document the practical value of the proposed approach with an empirical application, contributing to a never-ending debate for macroeconomists and policymakers: the macroeconomic effect and the transmission of monetary policy. 

We select several instruments recently proposed in the literature, specifically by \cite{gertler2015monetary} (GK), \cite{romer2004} (RR), \cite{miranda2021transmission} (MAR), the raw high-frequency surprise series of \cite{jarocinski2020deconstructing} (JK) and \cite{bauer2022} (BS) and compare the results between the estimates obtained from different model specifications using a Proxy SVAR with the unique set of responses estimated using a Proxy DFM. The choice of instruments is not casual. In fact, we consider also instruments that tend to capture the monetary policy shock along with a \textit{news component} coming from the central bank’s assessment of the economic outlook, which makes the underlying shock also partially anticipated \citep{ramey2016macroeconomic, jarocinski2020deconstructing, miranda2021transmission}. This event potentially exposes the VAR estimates to the problems introduced above. We find that the SVAR responses exhibit both output and price puzzles in almost all the specifications analyzed and across all instruments except the one proposed by MAR. On the other hand, the unique set of responses estimated through the Proxy DFM does not suffer from either of these puzzles. In other words, one can solve issues arising from the instrument by simply enlarging the information set of the econometrician, mitigating the effect of the \textit{news} component of the monetary policy shock.

In the last part of our paper, we provide a thorough analysis of monetary policy shocks transmission. We now exclusively focus on Proxy DFM results, using GK as the baseline instrument. In addition, we compare the results with all the instruments previously introduced. We study the various channels through which a monetary policy shock propagates, analyzing both impulse responses and the variance decomposition. 

The first interesting result is that our proposed model estimate impulse responses are robust across all instruments under analysis. Turning to the analysis of the transmission, we find that a monetary policy tightening shock is unequivocally contractionary for the economy. Real variables contract, but only after some months, with the exception of consumption and retail sales, which react at impact. All measures of prices decline, and their behavior indicates the presence of price rigidities in the economy, as they do not fully adjust at impact. The housing sector is deeply affected by the shock and, by weakening household balance sheets, can explain the sharp contraction in consumption \citep[see][for this mechanism]{mian2013household}. Its effect could compound with other important channels. For instance, equity prices fall, further negatively affecting household wealth; the exchange rate rises, signaling an appreciation of the domestic currency that translates into a reduction in net exports; and interest rates and spreads all point to an overall contraction of the financial sector, which could amplify the negative business cycle impact of the shock \citep{jorda2017macrofinancial}. The variance decomposition analysis indicates that monetary policy shocks are important in explaining business cycle fluctuations of the economy, with results that differ depending on the nature of the variable. For real variables, the shock does not explain much variance at the very impact, but its importance increases towards the third year. As expected, monetary policy shocks explain a large part of the variance already at impact for financial variables, exchange rates, and prices.

\paragraph{Related Literature.} 
Our study is linked to different strands of literature. First, we refer to the literature related to Dynamic Factor Models. This model can include a large number of variables in the estimation, and can therefore enlarge the information set of the econometrician, a feature which makes it useful both in forecasting and structural analysis \citep{forni2000generalized,stock2002forecasting}. On the latter, the model can be particularly appealing, given that it has been proved that the measurement error and non-fundamentalness issues are solved by construction. Furthermore, the estimation of its Wold representation - and reduced form shocks - does not depart too much from the standard procedure which is applied in the context of VAR models \citep{forni2009opening}. For this reason, the DFM has been widely used in the context of structural analysis in recent years, applied with different identification techniques - see, for instance, \cite{del200799,forni2010dynamic,barigozzi2014euro,luciani2015monetary,bjornland2019commodity,kerssenfischer2019puzzling,brignone2022evidence}. Among these, \cite{forni2010dynamic} is particularly related to our work as it was the first to explore the effects of U.S. monetary policy shocks in a DFM. In this paper we departure by using external instruments to identify the monetary policy shock rather than imposing timing restrictions of the variables in the model. Recent contributions related to our work, who analyze monetary policy shocks in a Proxy DFM are \cite{alessi2019response} and \cite{corsetti2022one}. The former uses it to estimate the response of asset prices to monetary policy shocks in the U.S., while the latter to study the heterogeneity in the transmission of shocks across euro area countries. Our work differs from theirs in terms of both methodology and scope. In a simulation exercise, we provide a comprehensive analysis of the advantages that a researcher can gain by using the Proxy DFM relative to the Proxy SVAR. In the empirical application, instead, we study the macroeconomic effects of monetary policy in the U.S., along with the channels through which it propagates.\footnote{Focusing on the issue of information and shock recoverability, our study is also related to the recent literature that has proposed milder conditions based on the concept of information sufficiency, see \cite{forni2019structural,chahrour2022recoverability}} Another major departure from them is our unit-variance shock estimation, which allows us to perform variance and historical decompositions.\footnote{In the paper, we only show the former decomposition.}

Second, our study focuses on the strand of literature which identifies structural shocks of interest in a VAR framework using external instruments available \citep[see][for a survey]{stock2018identification}. This approach typically involves a two-step strategy. Firstly, reduced form shocks are estimated, usually from a VAR model. Then, the structural shock of interest is identified by projecting a single selected reduced-form shock on the available external instrument. Its first appearance in macro-econometrics dates back to \cite{stock2008s}, reaching broader consensus some years later. Among the many contributions using Proxy SVARs we find \cite{stock2012disentangling}, which examine the macroeconomic dynamics of the Great Recession in the U.S. and the subsequent slow recovery, \cite{mertens2013dynamic}, which provide evidence on the U.S. personal and corporate income tax changes. More closely related papers are the ones using a Proxy SVAR to estimate the macroeconomic effects of monetary policy shocks. Among those, we find \cite{gertler2015monetary}, \cite{jarocinski2020deconstructing}, \cite{miranda2021transmission}, and \cite{bauer2022}. This literature has also been favoured by the development of the high-frequency identification (HFI) strategy, which allows using information from ``outside'' the VAR to identify the shock of interest.\footnote{The narrative approach, instead, has been less involved in the shock identification of Proxy SVAR. One potential reason is that it involves constructing a series for the shock from related historical documents, exposing this methodology to the problem of subjective choices.} The HFI exploits high-frequency asset price changes around policy meetings to quantify exogenous changes in monetary policy actions \citep[][among others]{kuttner2001monetary,gurkaynak2004actions}. The identifying assumption underlying the HFI approach is that unexpected changes in interest rates in a short window surrounding policy meetings are only due to monetary policy actions. 

In between the two bodies of literature mentioned above, our work is also related to \cite{miescu2019proxy,bruns2021proxy,de2023factor}, which use external instruments within a FAVAR approach. As we show in the simulations, FAVAR is able to address the \textit{missing variable} problem, but its estimates are still affected by measurement error and by model specification.

Finally, we also refer to the literature which directly focuses on the role of information in the proxy SVAR identification \citep[see the recent works by][]{forni2022external,plagborg2022instrumental,bruns2021proxy,miescu2019proxy}, and generally to the literature which has proposed milder conditions based on the concepts of informational sufficiency and \textit{recoverability} of the shocks\footnote{Recoverability requires that the structural shock is a linear combination of present and future values of VAR residuals. For this, fundamentalness implies recoverability, but not vice versa. } - see \cite{forni2019structural,forni2022external,chahrour2022recoverability}. The Proxy DFM is invertible by construction and, thus, the shocks are always recoverable, as the invertibility condition is stricter than the latter.

The remainder of the paper is organised as follows. Section \ref{sec:econometrics} describes the econometrics behind the DFM and the methodology we follow to apply the external instrument identification within this framework. Section \ref{sec:simulation} is devoted to the comparison between the Proxy SVAR and Proxy DFM results using a theoretical model with perfect foresight. Section \ref{sec:empirical} covers the empirical application to monetary policy. First, we compare impulse responses between the VAR and DFM identified with multiple instruments, then we use a Proxy DFM to shed light on the transmission mechanism of a monetary policy shock. Section \ref{sec:conclusion} concludes.

\section{Econometric framework}
\label{sec:econometrics}
In this section, we present the model and the identification procedure. Regarding the  DFM, we present two specifications. First, we show the stationary case, which is used in the simulation, which builds on \cite{forni2009opening}. Second, we present the methodology proposed by \cite{barigozzi2021large} to handle non-stationary variables. The latter is used in the empirical application. 

\subsection{Dynamic Factor Model}

\subsubsection{The stationary I(0) specification}
\label{sec:statDFM}

Consider a $N$-vector $x_t$ of weakly stationary time series. As standard in DFM literature, we assume each variable $x_{it}$, $i = 1,...,N$, can be rewritten as the sum of an \textit{idiosyncratic} component, $ \xi_{it}$, and a \textit{common} component, $\chi_{it}$. The former represents the source of variation affecting a specific variable and thus interpreted as measurement error or sectoral shocks. It is assumed that the $\xi_{it}$ are poorly correlated in the cross-sectional dimension, a milder and more realistic assumption than uncorrelation. This assumption is crucial for ascribing this model to the class of \textit{approximate} factor models, rather than the more traditional \textit{exact} factor models à la \cite{sargent1977business} and \cite{geweke1977dynamic}. On the other hand, the \textit{common} components, $\chi_{it}$, are assumed to permeate the entire dataset and to be function of $q$ common shocks $u_t=(u_{1t},u_{2t},...,u_{qt})'$, with $q<<N$, such that\footnote{In order to be consistent with the literature, throughout the paper, we will refer to $u_t$ interchangeably as common shocks or \textit{dynamic} factors.}
\begin{align}
	\chi_{it} &= b_{i1}(L)u_{1t} + b_{i2}(L)u_{2t} + ...+ b_{iq}(L)u_{qt} \nonumber 
\end{align}

Defining $\chi_t = (\chi_{1t} \dots \chi_{Nt})'$ and $\xi_t = (\xi_{1t} \dots \xi_{Nt})'$ we can rewrite the model in vector notation as
\begin{align}
	\chi_t &= B_\chi(L) {u_t} 
\end{align}

where $B_\chi(L)$ is a $ N \times q $ matrix, whose $(i,j)$-th entry is $b_{ij}(L)$, and $u_t$ is an orthonormal white noise vector such that $u_t \perp \xi_t$. Since the vector $u_t$ is orthogonal to  $\xi_{t}$, the latter is also orthogonal to $\chi_t$.  Moreover, being the vector $\chi_t$ singular, the dynamic factors are fundamental.

One can further rewrite the expression above in terms of \textit{static} factors as
\begin{align}
	\chi_t &= \Lambda F_t
\end{align}

where $F_t$ is a vector of $r>q$ static factors, still orthogonal to $\xi_t$, and $\Lambda$ is a $N \times r$ matrix of factor loadings. Notice that the static factors are only loaded contemporaneously, whereas we consider present and past values of the dynamic factors. It is further assumed that the vector $F_t$ follows a VAR of order \textit{p}, and that $F_t$ and $u_t$ are linked as follows
\begin{equation}
	D(L)F_t = \varepsilon_t, \quad \text{with  } \varepsilon_t = R u_t \label{chi} 
\end{equation}

where $D(L)$ is a $r\times r$ polynomial matrix of coefficients, $\varepsilon_t$ is the vector of VAR errors of the static factors, \textit{R} is a $r \times q$ matrix resulting from a spectral decomposition of the errors $\varepsilon_t$. Given that $r>q$, the stochastic vector $F_t$ is still singular. \par
By inverting the matrix of coefficient $D(L)$ we obtain the moving average representation 
\begin{equation}
	F_t = D(L)^{-1}Ru_t = B_F(L) u_t
\end{equation}
and the following relation
\begin{align}
	\chi_t &= \Lambda F_t = \Lambda B_F(L) u_t = B_{\chi}(L) u_t \label{chi_struct}
\end{align}

where $B_F(L)$ is a $ r \times q $ polynomial matrix of coefficients. Equations (\ref{chi}) - (\ref{chi_struct}) uncover the link between dynamic and static factors.	

\subsubsection{The non-stationary I(1) specification}
\label{sec:nonstatDFM}

In case the data is $I(1)$, we need a few adjustments, as suggested by the recent work of \cite{barigozzi2021large}. Let us suppose we have a $N$-vector $x_t$ of non-stationary time series. Considering a deterministic trend, one can still describe those as the sum of two orthogonal unobservable components, namely a common and idiosyncratic component, as
\begin{align}
	x_t &= \alpha + \beta t + \chi_t + \xi_t 
\end{align}
where $\alpha$ is a vector of constants and $\beta t$ is the linear trend. In this framework, moreover, the factors are assumed to be $I(1)$ and the idiosyncratic components are either $I(0)$ or $I(1)$.

The procedure in this case is as follows: (i) we estimate the number of static factors $r$ and dynamic factors $q$ on the $I(0)$ transformed data; (ii) we estimate the loadings $\Lambda$ from the differenciated data by means of the principal component; (iii) given $\Lambda$, we retrieve $F_t$ from the non-stationary dataset; (iv) we estimate $\alpha$ and the coefficient $\beta$ associated to the trend and we project $\hat{\tilde{x}}_t = x_t - \hat{\alpha} - \hat{\beta} t $ on the previously estimated loadings; (v) having $F_t$, we derive the MA representation of $\chi_t$, analogously to equation (\ref{chi_struct}), either applying a VAR-in-level specification or a VECM. The latter case requires the estimation of the number of cointegration relationships $d$ .

\subsection{External Instrument Identification in a DFM environment}
\label{sec:ident}

To identify the structural shocks from the estimated reduced form shock in (\ref{chi}), we follow the \textit{external instrument} procedure developed by \cite{stock2012disentangling} and \cite{mertens2013dynamic}. We start by describing the standard procedure in a quadratic system such as in VAR models, given that most of the assumptions also apply in a DFM framework. 

Let us assume we observe a variable $z_t$ satisfying the following conditions:
\begin{align}
	& \mathbb{E}_t(z_t \eta_{it}) = \alpha \label{condA}\\
	& \mathbb{E}_t(z_t \eta_{-it}) = 0 \label{condB}
\end{align}

with $\eta_{it}$ representing the structural shock we want to identiy and $\eta_{-it}$ the remainig structural shocks. The two above conditions state that the variable $z_t$ needs a) to be correlated with the structural shock $\eta_{it}$ we want to estimate and b) to be orthogonal to all the remaining structural shocks.

If the two condition apply, one can retrieve the structural shock $\eta_{it}$ from the estimated reduced-form shock $u_{it}$ by regressing $u_{it}$ on the instrument $z_{t}$ and then by a proper rescaling of  the coefficient coming from the regression. This, however, makes automatically impossible to retrieve a unit-variance shock and to perform variance and historical decomposition. Moreover, this way of proceeding would not fit in a DFM framework. Indeed, the methodology would require the choice of the estimated reduced-form residual which is directly associated to the policy variable. This is always possible in a VAR setting, given that it follows a non-singular quadratic $N \times N$ model, whereas the DFM is characterised by a singular $N \times q$ representation, which makes a clear one-to-one link between variables and shocks impossible. 

A viable solution is offered by \cite{forni2022external}, where they describe an alternative way to estimate a unit-variance shock using an external instrument identification in a non-singular quadratic system as the VAR.\footnote{A similar procedure is employed by \cite{alessi2019response}, which, however, is unable to retrieve the unit-variance shock and, therefore, to perform a variance and historical decomposition analysis.} We borrow from them and consider the projection of the instrument, $z_t$, on the vector of residuals, $u_t$, as follows
\begin{equation}
	z_t = \delta' u_t + v_t
\end{equation}

The unit-variance shock and the respective IRFs can be estimated as 
\begin{align}
	\hat{\eta}_{it} = & \frac{\hat{\delta}' \hat{u}_t}{\sqrt{\hat{\delta}' \widehat{\Sigma}_u \hat{\delta}}} \\
	C_{\chi,i}(L)  = & B_{\chi}(L) \widehat{\Sigma}_u \frac{\hat{\delta}}{\sqrt{\hat{\delta}' \widehat{\Sigma}_u \hat{\delta}}}
\end{align}

where $\Sigma_u$ is the variance-covariance matrix of $u_t$, which is equal to the identity matrix in our framework.\footnote{Additional calculations are available in Section \ref{sec:unitvarshock} of the Appendix.} We obtain that the instrument $z_t$ is correlated to the structural shock of interest $\eta_{it}$, spanned throughout all the $u_t$, and that it is orthogonal to all the remaining structural shocks $\eta_{-it}$. As a direct consequence, we do not need to choose a specific variable to instrument, as is the case in standard VARs - a choice that has often left room for discussion in the literature, providing an additional source of sensitivity that could reduce the external validity of the results.\footnote{One recent example is \cite{bauer2022}, which argue that the two-year Treasury yield is a better measure of the monetary policy stance than the one-year Treasury yield used by \cite{gertler2015monetary}.}

\section{ Non-fundamentalness and Proxy identification}
\label{sec:simulation}
This section aims at showing some of the possible advantages offered by the proxy identification strategy in the DFM framework, as described in section \ref{sec:ident}, compared to a standard VAR model. We conduct the first part of the analysis by deploying a theoretical macroeconomic model. Specifically, we borrow the simple Real Business Cycle (RBC) model of \cite{leeper2013fiscal}. Such model is characterized by log preferences, inelastic labor supply, full capital depreciation and fiscal foresight. The latter, in particular, is what matters in our case. In fact, in the presence of fiscal foresight, agents are able to foresee future values of the tax rate and behave according to such information.

The evolution of capital through time is dictated by the following equation 

\begin{equation}
	k_t=\alpha k_{t-1}+a_t-\kappa \sum_{i=0}^{\infty} \theta^i E_t \tau_{t+i+1} \label{capital}
\end{equation}

where $0<\alpha<1, |\theta| <1, \kappa = (1-\theta)\tau/(1-\tau)$, $\tau$ is the steady state tax rate, $k_t$ is capital, $a_t$ is technology and $\tau_t$ is tax rate. All the variables are considered as log deviations from the their steady state value.

For the sake of simplicity, technology and tax are assumed to follow two \textit{i.i.d.} processes, $u_{a,t}$ and $u_{\tau,t}$, respectively.  In order to allow for fiscal foresight, it is assumed that 

$$ \tau_t = u_{\tau,t-h}$$  

Depending on the value of $h$, we face different scenarios. When $h=0$, equation (\ref{capital}) reduces to $ k_t=\alpha k_{t-1}+a_t $, $i.e.$  the representative agent does not have information about the future. As a consequence capital accumulation does not depend on the tax shock. In this situation, the information set of the econometrician is aligned with that of the agent. A completely different case, instead, is when $h>0$. Since the information set at time $t$ of the agent includes present and past observations of $u_{a,t}$ and $u_{\tau,t}$, she has knowledge about future values of $\tau_t$, according to which capital will be adjusted. Such knowledge generates misalignment between the two information sets, causing non-fundamentalness of the shocks.

In what follows, we consider the case with $h=2$, that is characterized by the following equations
\begin{align}
	a_t &= u_{a,t}  \nonumber \\
	k_t &= \alpha k_{t-1} + a_t -\kappa (u_{\tau,t-1} + \theta u_{\tau,t} ) \nonumber\\
	\tau_t &= u_{\tau,t-2} \nonumber
\end{align}

which implies the MA representation

\begin{equation}
	\left(\begin{array}{l}
		a_t \\
		k_t \\
		\tau_t
	\end{array}\right)=\left(\begin{array}{cc}
		0 & 1 \\
		\frac{-\kappa(L+\theta)}{1-\alpha L} & \frac{1}{1-\alpha L} \\
		L^2 & 0
	\end{array}\right)\left(\begin{array}{c}
		u_{\tau, t} \\
		u_{a, t}
	\end{array}\right)=B(L) u_t . \label{MAff}
\end{equation}

As pointed out by \cite{forni2020common}, $u_t$ is fundamental for none of the square subsystems generated by the possible pairs of $ y_t = (a_t,k_t,\tau_t)'$. On the other hand, considering all the three variables is sufficient to recover $u_t$.\footnote{The fact that adding information helps in recovering the true structural shocks is not new in the literature. See \textit{inter alia} \cite{giannone2006does}.}

\subsection{Simulation}

We propose a simulation exercise to quantitatively assess to what extent non-funtamentalness threatens the results coming from the empirical analysis. The exercise is very close in spirit to the one performed in \cite{forni2020common} and \cite{miescu2019proxy}.

We consider a static factor model representation of the form 

\begin{equation}
	\chi_t = \Lambda F_t \label{static}
\end{equation}

where $\chi_t$ is a $n$-dimensional vector of economic variables, $\Lambda$ is a $n \times r$ matrix of loadings and $F_t$ is a vector of static factors of dimension $r$.

Let us define $ F_t = (k_t, u_{a, t}, u_{\tau, t} u_{\tau, t-1}, u_{\tau, t-2})'$, which is assumed to have the following VAR(1) dynamics

\begin{equation}
	F_t = A F_{t-1} + B u_t
\end{equation}

where 
\begin{equation*}
	A = \left(\begin{array}{ccccc}
		\alpha & 0 & -\kappa & 0 & 0 \\
		0 & 0 & 0 & 0 & 0 \\
		0 & 0 & 0 & 0 & 0 \\
		0 & 0 & 1 & 0 & 0 \\
		0 & 0 & 0 & 1 & 0
	\end{array}\right) \quad \quad 
	B = \left(\begin{array}{cc}
		1 & -\kappa \theta \\
		1 & 0 \\
		0 & 1 \\
		0 & 0 \\
		0 & 0
	\end{array}\right) \quad \quad
	u_t = \left(\begin{array}{c}
		u_{a, t} \\
		u_{\tau, t}
	\end{array}\right)
\end{equation*}

Having the factors $F_t$, one can easily find that $y_t = \bar\Lambda^x F_t$, where $ y_t = (a_t,k_t,\tau_t)'$ and 

\begin{equation*}
	\bar\Lambda^x = \left(\begin{array}{ccccc}
		0 & 1 & 0 & 0 & 0 \\
		1 & 0 & 0 & 0 & 0 \\
		0 & 0 & 0 & 0 & 1 
	\end{array}\right) 
\end{equation*}

We assume the econometrician observes $ \chi_t = (x_t', x_t^{*} {'})'$ where $x_t = (\tau_t, k_t)'$ is a non-fundamental subsystem of $y_t$ and $x^*_t$ is a set of $n$ survey series generated artificially. The cross-sectional dimension of the entire sample is thus $N = n + 2$. The series in $ x^{*}_t  $ are generated by a linear transformation of the factors, given by $\Lambda^* F_t $, where the entries of $\Lambda^*$ are drawn from independent $N(0,1)$. 

Considering $\Lambda^x$ as the third and second rows of $\bar\Lambda^x$ in such order, we can rewrite equation (\ref{static}) as

\begin{equation}
	\left(\begin{array}{c}
		x_t  \\
		x^{*}_t 
	\end{array}\right) = \left(\begin{array}{c}
		\Lambda^x  \\
		\Lambda^* 
	\end{array}\right) F_t
\end{equation}

Finally, we assume that the varibles are observed with a measurement error
\begin{equation}
	Y_t = \chi_t + \xi_t  = \Lambda F_t + \xi_t 
\end{equation}

where $\xi_{it} \sim N(0,\sigma_i) $ and $\sigma_i \sim U(0,\nu) $, with $\nu \in [0.5, 2, 5]$. To higher values of $\nu$ corresponds an higher share of measurement errors. 

We simulate $1000$ different datasets of $T = 200$ observations from the model presented above, using the parametrization of \cite{leeper2013fiscal}: $ \alpha = 0.36, \theta = 0.2673, \tau = 0.25$ and $u_t \sim N(0,I)$. We arbitrarily fix $n=100$, so to have a sufficiently large cross-section. 

For each dataset, we estimate a standard VAR, a DFM and, in addition, the Factor Augmented VAR (FAVAR) of \cite{bernanke2005measuring}.\footnote{ The inclusion of the FAVAR is motivated by the work of \cite{miescu2019proxy}, that proposed a Proxy FAVAR, which is clearly related to our methodology. Differently, their simulation is based on an enlarged version of the theoretical model of  \cite{leeper2013fiscal}, but the results are coherent to ours. Our choice of a restricted number of variables is due to the sake of simplicity.} Unless otherwise stated, in each case we set the lag order equal to $2$. \footnote{ We repeated the estimation considering more lags and the findings are virtually identical. Results are available from the authors upon request.}

While DFM and FAVAR estimation involves the whole sample, the VAR analysis includes only few key variables, specifically capital and tax rate, though we add also technology in some exercises. Moreover, the DFM is estimated with parameters $r = 5$ and $q = 2$, whereas the FAVAR is based on a bivariate VAR augmented with three additional factors.\footnote{In this way we have that the space generated is the same as the one generated by the $F_t$ of the DFM}.

To identify the shock from the simulated series for all the three models, we apply the external instrument procedure described in Section \ref{sec:ident}. As instrument,  we consider the structural shock itself, \textit{i.e.} $u_{\tau,t}$, which is the best instrument one can possibly have.\footnote{We also consider instruments of different quality. Since the results do not show substantially difference with those presented here, we show them as robustness in Appendix \ref{sec:dirtyinsrtu}.} Furthermore, we repeat each exercise with different and increasing values of $\nu$, so that we are also able to observe the role of the measurement error. 

\begin{figure}[h!]
	\caption{Comparison of VAR specifications}  
	\includegraphics[trim=4cm 0cm 3.5cm 0cm,clip=true,width=1\textwidth]{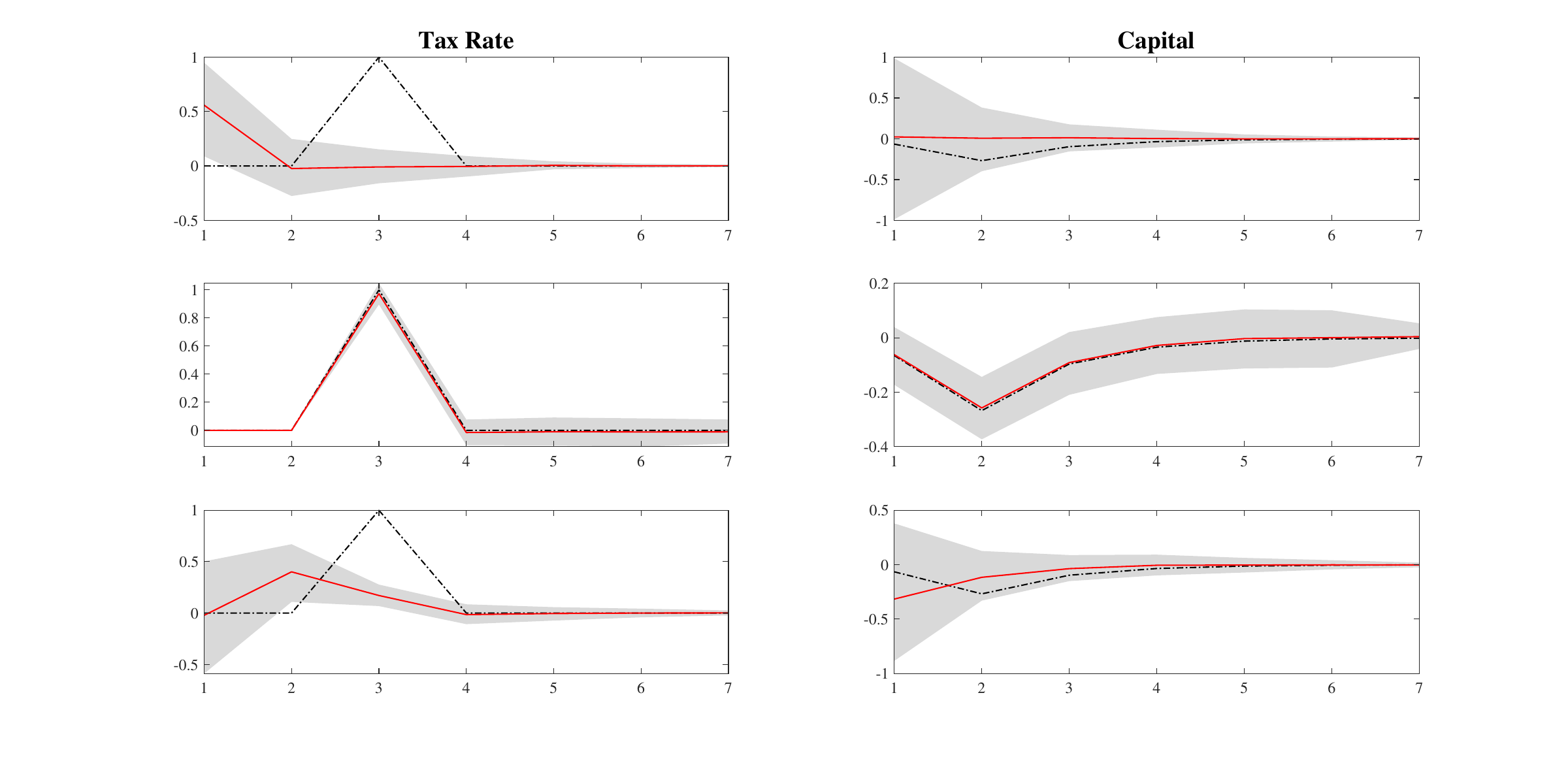}
	\subcaption*{Notes: black dotted lines are the theoretical IRF, the red lines are the responses obtained estmating a VAR with different specifications along with their $68\%$ confidence bands interval in grey. The empirical IRF are computed as the sample average of the resposes obtained across simulations. Going form the upper panel to the lower, we see the results for a bivariate VAR(3) on capital and tax rate observed without measurement errors, for a trivariate VAR(3) on capital, tax rate and technology without measurement error and finally the same trivare VAR(3) obeserve with small measuremnt error , specifically $\nu = 0.5$.}
	\label{VARres}
\end{figure}

\subsection{Simulation results}

In the next section we show the possible issues that may affect the IRFs estimated with a proxy identification within a VAR framework. 

Figure \ref{VARres} plots the responses of capital and tax rate to a tax shock, for three different VAR specifications. In each case, the series are simulated with two-periods fiscal foresight and the IRFs are computed as the average across the model simulations.
The black dotted lines in the figure represents the model true responses. 
The agents know two periods in advance how and when a tax shock will hit, therefore they adjust capital accordingly. However, when the same responses are estimated from the data simulated from the same model, the econometrician is not always able to recover the true responses. In fact, the results depend on two different aspects: first, on the information set at the econometrician's disposal and, second, on whether what is observed is contaminated by the measurement error.

The results reported in the upper panels of Figure \ref{VARres} are derived from a bivariate VAR on capital and tax rate, where no measurement error is added. Clearly, the VAR model is unable to recover the structural shock, even without measurement error contaminating the estimation, thus providing misleading impulse responses. Indeed, as explained in the previous section and more in detail in \cite{leeper2013fiscal}, considering only capital and tax rate would be insufficient to solve the non-fundamentalness issue, leading to misspecifications. 
In the middle panel we show that adding a third variable, namely technology, is crucial for the reliability of VAR estimates. The additional variable allows VAR empirical responses to retrace almost perfectly their theoretical counterparts.
Nevertheless, in reality, variables are often observed with measurement errors. We then augment the same trivariate system with a small measurement error (obtained setting $\nu = 0.5$), and proceed in identical manner. The results, shown in the last panel of Figure \ref{VARres}, highlight that the overall estimates of the true IRFs can still be distorted by measurement errors, despite the system itself is fundamental.

\begin{figure}[h!]

	\caption{IRF of tax rate and capital to a tax shock, with a two-period fiscal foresight.}
	\includegraphics[trim=3.8cm 0cm 3.5cm 0cm,clip=true,width=1\textwidth]{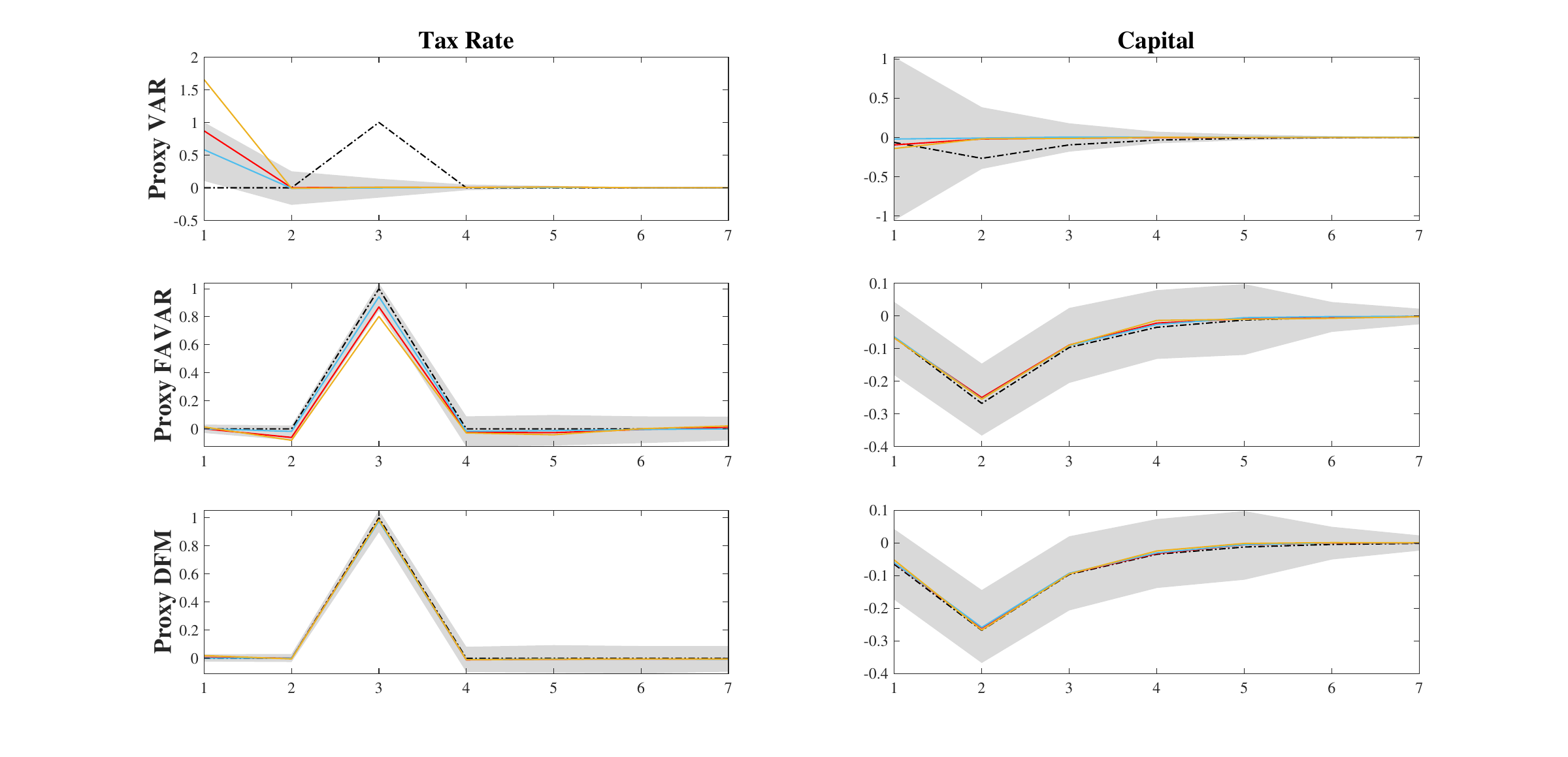} 
	\subcaption*{Notes: Black dotted lines are the "true" responses. Responses obtained in case $\nu = 0.5$ are displayed in blue, along with the $68\%$ confidence bands in grey. The red lines are the response obtained in case $\nu = 2$. The yellow lines are the response obtained in case $\nu = 5$. In each case the external intrument is the real structural shock $u_{\tau,t}$. The empirical IRF are computed as the sample average of the resposes obtained across simulations. Going form the upper panels to the lower, we see the results of the bivariate VAR(2) on capital and tax rate, of the FAVAR(2) on capital, tax rate and three factors and finally of the DFM(2).}
	\label{simuclean}
\end{figure}

Therefore, the estimated responses following a proxy identification procedure can in principle be affected by both non-fundamentalness and the measurement error issues. As a solution, we propose to apply the same identification procedure into a DFM frameowork.
The comparison is given in Figure \ref{simuclean}, which complements the results displayed in the previous chart. In this case, we still have a model with a two-period fiscal foresight, with the true impulse responses of the tax rate and capital to a tax shock which are depicted in black dotted line.
We add different levels of measurement error to the variables - specifically considering $\nu=0.5, 2$ and $5$, and we then re-estimate the shock following the same proxy identification technique for each level of  $\nu$. The estimated IRFs are reported in blue, red and yellow for each level of the added measurement error.

Moreover, in figure \ref{simuclean} we compare the results coming from three different models: bivariate VAR (top panel), FAVAR (middle panel) and DFM (bottom panel). Not surprisingly, the higher the measurement error, the more distorted are the estimated responses of the VAR. Conversely, both the FAVAR and DFM successfully recover the true responses, although we observe an increase in the estimation-bias in the former as the error increases. 
These results underline the usefulness of expanding the information set at econometrician's disposal also in the proxy identification procedure: not only it helps solving the non-fundamentalness issue by construction, but one can also better deal with the measurement error problem, which usually affects macroeconomic time series. Specifically on this matter, the DFM behaves better then the FAVAR because in the latter some of the variables are observed with the measurement error, a feature which can lead to a contamination of the estimated IRFs, as we will observe later in this section.

Figure \ref{shocks_by_models} complements the above picture, though focusing this time on the underlying shock which is estimated by the different models. The chart is obtained following the same procedure explained above for the IRFs, and it depicts the true shock simulated from the model (black dotted line) along with the unit-variance shocks obtained from a VAR (top panel), FAVAR (middle panel) and DFM (bottom panel) which are estimated each time with different level of measurement errors.\footnote{As for the IRFs, the chart plots the mean of the distribution of shocks which are estimated for each model simulation} 
As evident, the non-fundamental VAR is never able to recover the true structural shock and higher measurement errors contribute to further distorting the estimated shock. On the other hand, FAVAR and DFM perform largely better, especially the latter which shows great robustness. 

In Table \ref{frob_full}, we also report the Frobenius Norm computed between the true model IRFs and structural shocks and the estimated counterparts.
Results corroborate what seen so far. Using VAR as benchmark, we show that the additional information exploited by FAVAR and DFM is effective. Let us focus on the specific case when the model is simulated with fiscal foresight: here both DFM and FAVAR largely overcome VAR. As $\nu$ increases their results tend to converge to those of VAR, but at different rates: DFM converges at a lower rate with respect to the FAVAR, suggesting a higher reliability of DFM result.

For a matter of completeness, we also briefly comment the results obtained following the same procedure, but with a model with \textit{no fiscal foresight}. In this model specification, the econometrician can estimate a bivariate VAR from of the simulated series of tax rate and capital to have a sufficient information set, \textit{i.e.} the VAR is fundamental.\footnote{See \cite{forni2014sufficient} for a discussion.} 
However, although results are quantitatively different from the case studied in the previous paragraphs, they still lead to analogous conclusions. We spend only few words describing the results of figure \ref{simuclean_NFF}, as it is not crucial in our analysis. Here we observe that the the VAR estimated on the simulated series with small measurement error (blu line) is now able to estimate reliable responses, which are almost identical to those of the other models. However, as measurement error increases (red and yellow lines), the VAR estimation deteriorates quickly, whereas the other two models are consistent. We refer again to table \ref{frob_full}, which help us complete the overall picture. By looking at the Frobenius norm, it is evident that with the measurement error increasing, both the DFM and FAVAR perform better with respect to the VAR. \footnote{ This result seems in contrast with what observed before under the case of fiscal foresight, where an increase of measurement error was pushing the factor models closer to VAR in terms of distance. A possible explanation lies upon the different rates of deterioration. If variables are only affected by measurement error, VAR deterioration rate is higher than those of the other two models. This is because the latter models are able to better handle the issue and manage to better recover theoretical results. Thus, when this error increases, DFM and FAVAR perform relatively better than VAR. Conversely, when there is fiscal foresight, VAR is informationally deficient and, consequently, is already largely distorted by non-fundamentalness. The distortion due to the increasing measurement error add up to the pre-existing one, but the sum deteriorate at a slower rate wrt DFM and FAVAR, reducing the performance gap. The conclusion we draw from the table is: VAR is largely affected by measurement error, but even more by informational deficiency. In other words, ift the former is of small size - and the shock to be estimated is fundamental - VAR may be comparable to DFM and FAVAR and its results are reliable. Conversely, if the shock is no-fundamental, but we do not observe any measurement error contaminating the series, the estimation will always be distorted.}

As a final note, we stress another point which is also important in the estimation of the IRFs, represented by the variables choice and the overall model specification. We do it through an exercise which differs from the ones we showed above, where we show that not only the VAR, but also the FAVAR is highly dependent on such choice. We proceed as follows: we estimate a trivariate VAR and a FAVAR having three variables observed with measurement errors plus two factors. We fix the first two observed variables in both the VAR and the FAVAR to be capital and tax rate, whereas the third variable changes for each iteration. The measurement error, instead, is always kept equal to $\nu= 0.5$. Figure \ref{modelspec} compares the $n$ different estimates of the two models, with the VAR shown in the top panel, while FAVAR in the middle one. As evident, it is sufficient to vary only one variable at the time to obtain responses that greatly differ across specifications. In some cases, the results may end up being very misleading. This happens because each observed variable brings in the model a different type of information, along with extra measurement error, which can therefore contaminate the results. Conversely, the DFM does not have this weakness by construction: the econometrician already has all the information set available at his disposal, and the variables are already cleaned by the measurement error.

\section{Empirical application}
\label{sec:empirical}

In this section, we provide a detailed account of our empirical analysis by describing the data and the model specification used. We present our findings in two sub-sections. Firstly, we compare impulse responses obtained using a Proxy VAR versus a Proxy DFM. For the analysis, we identify the monetary policy shock with a broad range of external instruments available in the literature. Then, we explore the transmission of monetary policy shocks in the United States, using the Proxy DFM identified with all the instruments previously analysed. Finally, we present the variance decomposition obtained using the new ``unit-variance shock" methodology, which we have described in section \ref{sec:ident}.

\subsection{Data, specifications and procedure}

We use data from the FRED monthly dataset, which is described in \cite{mccracken2016fred}, covering the period from January 1963 to December 2018. The dataset contains $N=100$ macroeconomic and financial variables. For the VAR model, we specify a \textit{core} subset of variables, which include the industrial production index, the unemployment rate, the consumer price index (CPI), and the policy rate. We use the one-year Treasury yield as the policy variable, as it is common in the literature. In contrast, the DFM includes all available variables in the dataset. The variables are left in levels or log-levels and are not transformed to reach stationarity.\footnote{It is important to note that we do not perform a VECM estimation in either the VAR or DFM cases. \cite{sims1980macroeconomics} and \cite{sims1990inference} show that the cointegration relationship is correctly taken into account within a standard VAR in levels, at least at short horizons. Similarly, as shown by \cite{barigozzi2021large}, the VAR estimated on $I(0)$ static factors produces IRFs that, at a short horizon, are equal to the VECM specification, without the need to explicitly estimate the number of cointegration relationships.}

We estimate a VAR with $p=8$ lags, chosen based on both the Akaike and Schwarz information criteria. These criteria suggest $p=11$ and $p=6$, respectively. The estimation of the DFM in based on Section \ref{sec:nonstatDFM}, and we determine the number of static and dynamic factors using the tests provided by \cite{bai2002determining} and \cite{hallin2007determining}, respectively. Based on these tests, we set the number of static factors to $\hat{r}=9$ and the number of dynamic factors to $\hat{q}=4$, which represent our baseline parameter specification of the model. To maintain consistency with the VAR, we set the number of lags to $p=8$. We conduct a robustness analysis, as presented in Appendix \ref{sec:robustness}, and find that the results are virtually identical when varying $p$, which holds true for both the VAR and DFM models.

After having estimated the models, we identify the structural shock of interest exploiting the external information provided by the proxy (see Section \ref{sec:ident}). The baseline analysis is based on GK, which covers a period stretching from January 1990 to June 2012. However, we also offer a comparison with instruments developed by RR, MAR, JK, and BS.\footnote{Regarding RR, we take the version extended by \cite{miranda2020us}.} As mentioned in Section \ref{sec:ident}, our new procedure has the advantage of being agnostic regarding the choice of the policy variable to instrument. Indeed, we do not need to choose between the one- and two-year Treasury yield to capture the monetary policy stance, as long as the information of the structural shock we want to estimate (in our case, the monetary policy shock) is spread throughout the $q$ estimated reduced-form innovations.

\subsection{A VAR-DFM comparison across instruments}

\begin{figure}[h!]
	\caption{VAR vs DFM: proxy comparison} 
	\centering
    \includegraphics[scale=0.3]{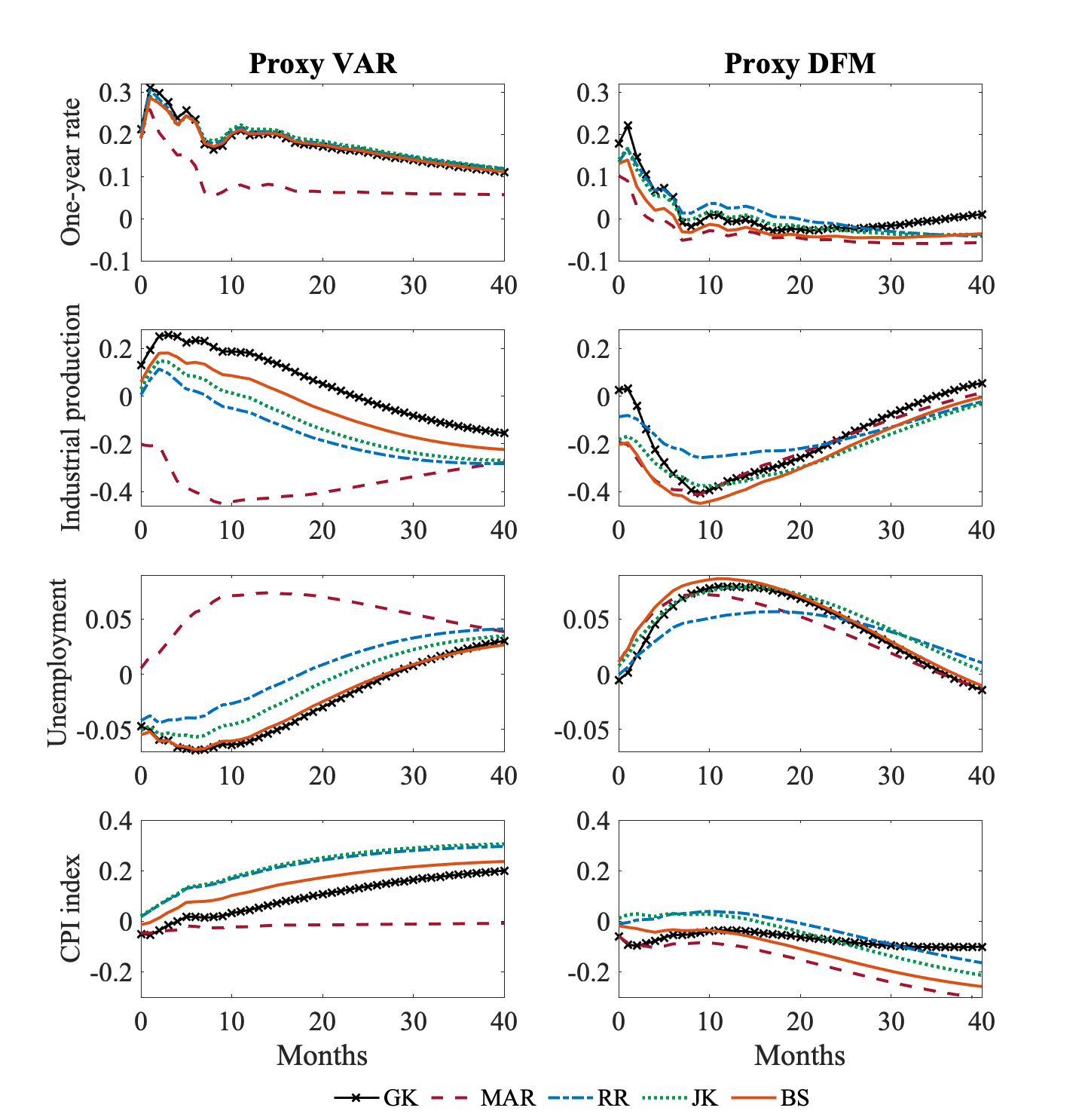}
	\subcaption*{Notes: Comparison between impulse responses identified with all instruments for monetary policy shocks included in the analysis. The VAR specification includes our \textit{core} variables: the one-year rate, industrial production, unemployment and CPI.}
	\label{Instr_comp}
\end{figure}

The aim of this section is to compare the estimated impulse responses from the \textit{core} specification of the VAR identified with the most common instruments in the literature, namely GK, RR, MAR, JK, and BS, with those of the DFM obtained with the same instruments. The comparison is shown in Figure \ref{Instr_comp}. Three main points can be made here: firstly, there are significant differences between the impulse responses estimated with the two models; secondly, in most cases, the VAR model produces results that are at odd with standard macroeconomic theory, with both price and output puzzles, whereas the DFM model does not; lastly, the VAR estimates show large variability across instruments, while by construction the DFM is unaffected by this issue.

Overall, the VAR estimates show that a contractionary shock raises industrial production and prices and lowers the unemployment rate. This issue is related to the ``information effect," i.e., the instrument not only contains information about the underlying monetary policy shock, but at the same time it also carries information, or \textit{news}, on the future macroeconomic outlook which is implicitly comunicated by the central banks when during the policy announcements. Therefore, the instrument is not exogenous and can be predicted with any publicly available information at the time of the FOMC announcement. The puzzles are evident for the responses obtained for all the instruments except with MAR. This is because the authors have developed an informationally-robust instrument that combines the high-frequency approach with the central bank's information set (Greenbook forecasts) to control for the macroeconomic information revealed by the central bank with its policy change.\footnote{The main assumption of this paper is that they assume that if the econometrician can correctly identify the monetary policy shock, then private agents should respond to it as a true monetary policy shock. The problem, however, is that private agents do not have the Greenbook forecasts in real time, as they are released to the public with a five-year lag.} This is not the only attempt in the literature, and others have addressed the same issue in different ways. \cite{jarocinski2020deconstructing} take a high-frequency approach and distinguish a true monetary policy shock from an information shock by looking at the comovement of interest rates and stock prices around policy announcements. \cite{bauer2022} obtain the series of high-frequency monetary policy surprises and then orthogonalize them using a set of predictors that are closely related to the Fed's monetary policy rule. Given our interest in exploring the role of our model in eliminating any kind of puzzle in the estimation of impulse responses without subjectively choosing how to \textit{clean} the instrument, we use the raw high-frequency surprise series for JK and BS.\footnote{The \textit{cleaned} version for MAR is simply GK.}

On the other side, performing the same identification strategy in a DFM environment implictly solves the information for those instruments which were affected. This is linked to the the large dimension of data which is at econometrician's disposal: including variables which also carry information about the future helps \textit{purging} the responses from the news component. Therefore, the estimated effect of monetary policy using the Proxy DFM yields puzzle-free impulse responses across all instruments: a contractionary shock that raises the one-year government bond yield lowers industrial production and prices (with some lag for some instruments), while raising the unemployment rate. Clearly, the use of a model that incorporates a large amount of information is critical to recover plausible monetary policy responses, no matter the instrument.

The problem of a limited information set available in small- and medium-scale models also determines a large variability in the estimates when including additional variables \citep[see][]{forni2020common}. We proceed along their lines and show the impulse response for the \textit{core} variables when one additional variable is added to the model at a time, with a total of 95 different specifications. We also test for each model specification whether the shock is invertible. To do this, we use the test recently proposed by \cite{forni2022external}, where the proxy of interest is projected on the current value and the first $r$ leads of the Wold residuals $v_t$ as follows:
\begin{equation} \label{eq:test_invert}
	z_t = \sum_{k=0}^{r}\hat{\gamma}_k^{'}\hat{v}_{t+k}+\hat{\xi}_{r,t}
\end{equation}

where $z_t$ is the proxy, and $v_t$ are the VAR reduced-form residuals. We test for invertibility using the F-test for the significance of the regressors, with the null hypothesis being $H_0:\gamma_1=\gamma_2=\dots=\gamma_r=0$ against the alternative that at least one of the coefficients is non-zero.\footnote{The regression does not include a constant as $E(z_t|x=0)=0$, where $x=\sum_{k=0}^{r}\hat{\gamma}_k^{'}\hat{v}_{t+k}$.} We estimate the regression in equation (\ref{eq:test_invert}) with $k=8$ and a 5\% confidence level. We choose this calibration because too many leads would introduce significant noise in the regression and too few would be insufficient, both of which would undermine the validity of the test. If the estimated VAR specification has a shock invertible, then the corresponding impulse response is colored in yellow. Grey lines, instead, represent models with non invertible shocks.

\begin{figure}[h!]
	\caption{VAR vs DFM with GK identification: IRFs comparison}\label{VAR-DFM1}
\hspace*{-2cm} \includegraphics[scale=0.29]{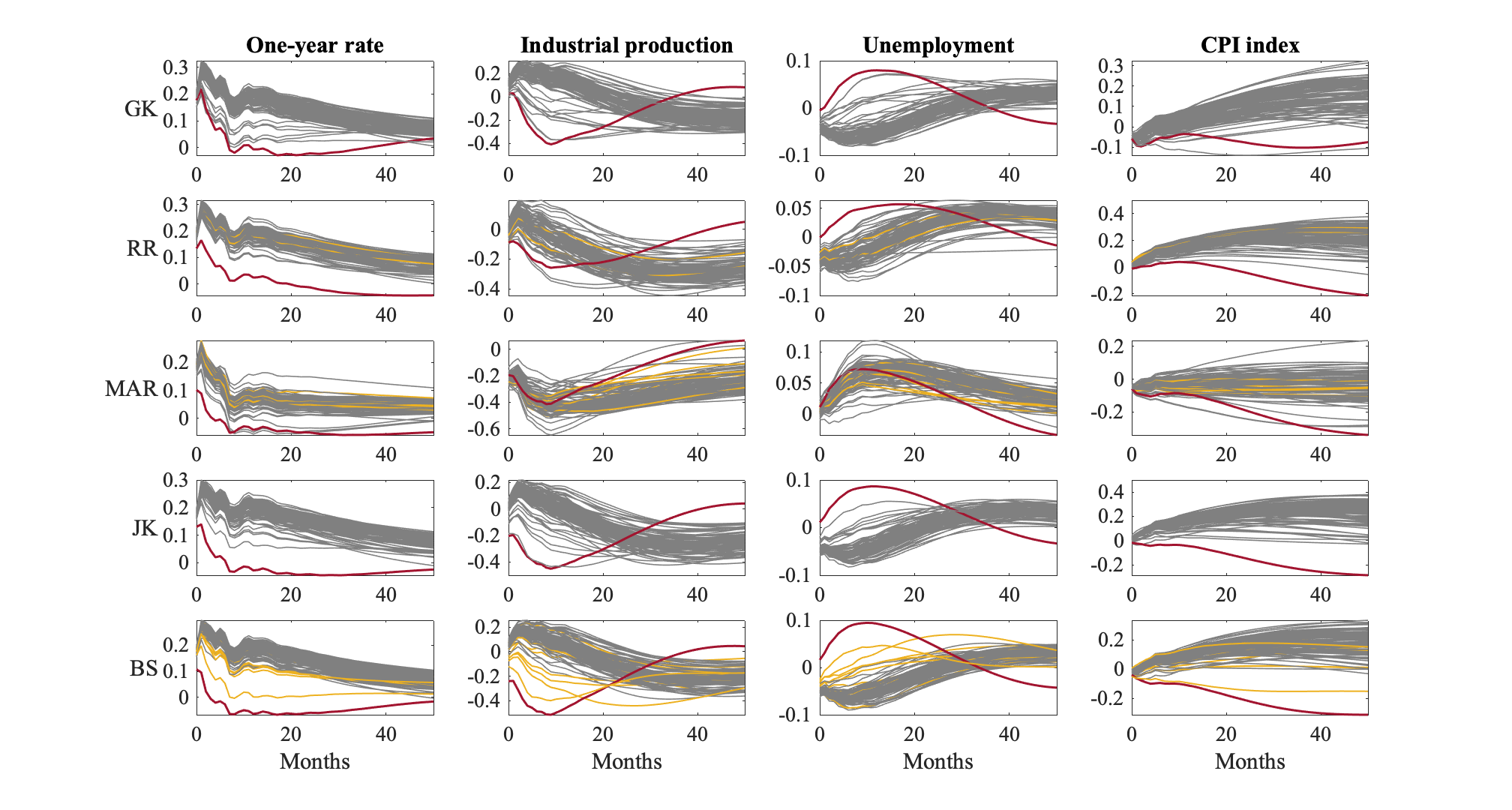}
	\subcaption*{Notes: The Figure shows impulse responses to a monetary policy shock estimated from the 95 different VAR specifications (grey lines=non invertible shock; yellow lines=invertible shock) and the DFM (red line). The identification is performed using all external instruments used in this paper. The invertibility test in equation (\ref{eq:test_invert}) includes 8 leads and uses 5\% as confidence level.}
		\label{fig:VAR_DFM_specifications}
\end{figure}

The results are plotted in Figure \ref{fig:VAR_DFM_specifications} and indicate that the additional variable added at each iteration to the \textit{core} specification is sufficient to include additional information, along with measurement error, that can significantly affect the estimated impulse responses. However, it is often not sufficient to overcome the puzzles in the estimates, and even when the impulse responses are estimated with a cleaned instrument (MAR), some specifications estimate price puzzles. Overall, the high sensitivity of these estimates underscores the importance of carefully selecting the variables to be included in the model.\footnote{This problem affects not only the impulse responses but also the underlying structural shocks.} An additional issue highlighted in the figure is that in most of the cases, the shock is not invertible (gray line). Conversely, the DFM specification does not suffer from the above problems because it is able to deal with a large number of variables simultaneously, which is not possible in the VAR framework due to the curse of dimensionality. Lastly, as described in section \ref{sec:statDFM}, the DFM by construction cleans the data from the measurement error, thereby eliminating the corresponding bias in the estimated responses.

\subsection{Propagation Channels of the Monetary Policy Shock}

This section further investigates the various channels of monetary policy \citep[see][for a survey]{mishkin1995symposium} and, at the same time, shows the power of the Proxy DFM which succesfully estimates robust puzzle-free impulse responses across all the analysed external instruments. Compared to VAR models, a distinct advantage of our methodology relies in the possibility to study a wider range of variables at the same time, allowing for a broader understanding of the monetary policy transmission mechanism. Using the newly developed ``unit-variance shock", we are also able to estimate the variance decomposition, which was not previously feasible within the external instrument approach (see section \ref{sec:ident}).

Figure \ref{MP_transmission} plots the impulse responses of a representative sample of variables selected to explore the transmission channels of monetary policy. Specifically, we look at measures of economic activity, labor, housing, financial and labor markets, exchange rates, and uncertainty. The figure reports the median impulse responses obtained from the GK instrument, along with the 68 and 95 percent confidence bands. However, the median responses using all of the other instruments analysed in the previous section are also reported. Notice that, for comparability reason, all the results are normalized at 100 basis point increase in the one-year Treasury yield at impact, a fairly common normalization in the literature. 

\subsubsection*{Real Economy and Labor Market}
\begin{figure}[h!]
	\centering
	\caption{The transmission of monetary policy}
	\hspace*{-2.65cm}	\includegraphics[scale=0.23]{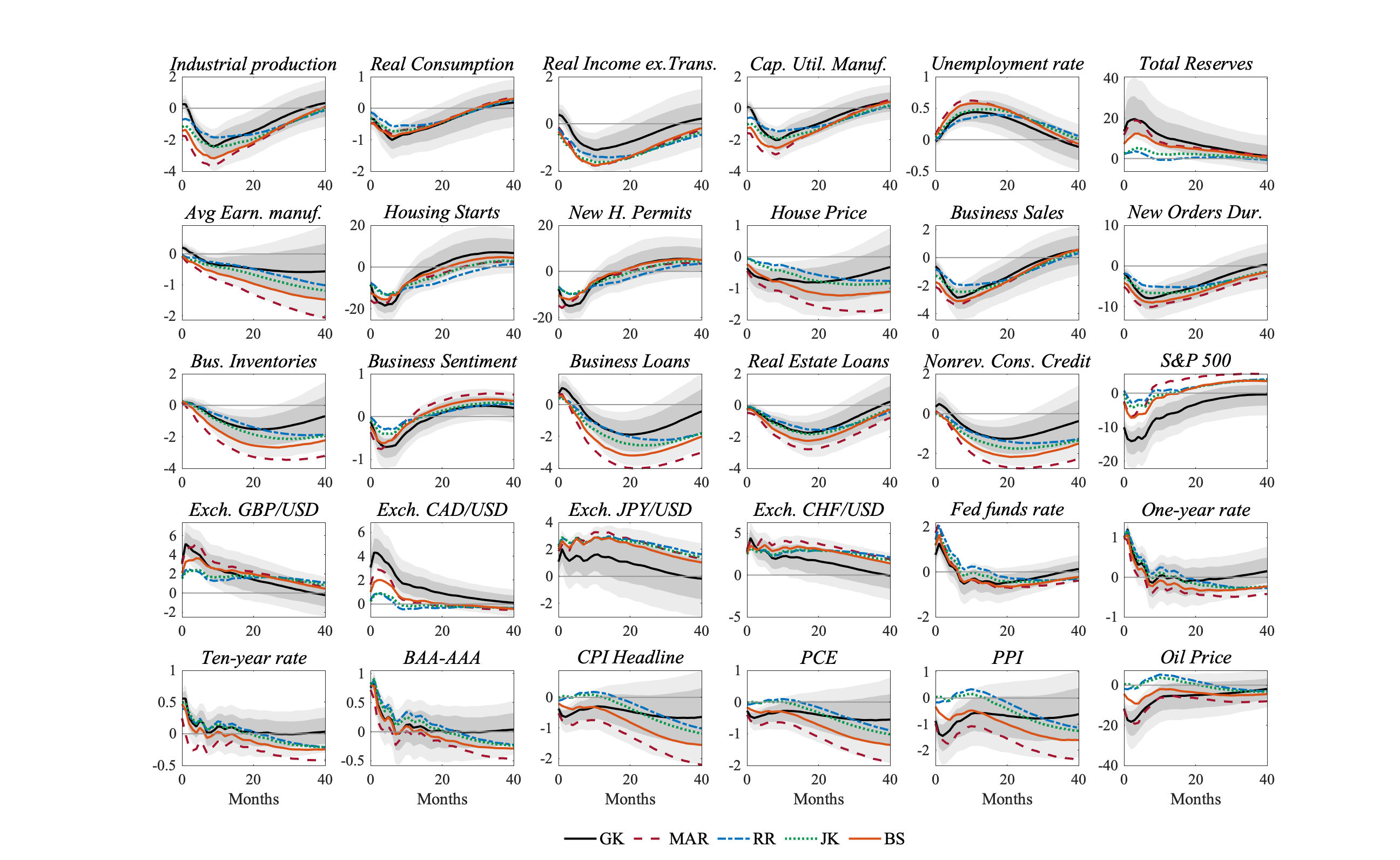}
	
	\subcaption*{Notes: Impulse response functions of a monetary policy shock identified using GK, MAR, RR, JK, and BS as instruments in a DFM framework. The impact of the shock is normalised such that the one-year Treasury yield increases by 100 basis points at impact. The reported confidence bands are related to GK only and correspond to the 84th (dark grey) and the 95th percentile (light grey) respectively, and are estimated via wild bootstrap.}
	
	\label{MP_transmission}
\end{figure}

Figure \ref{MP_transmission} shows that a contractionary monetary policy has a negative effect on the economy. Industrial production reacts with a delay of two months and builds up gradually over time to reach its peak impact after 10 months at -2\% for GK before returning to its original level, while the other instruments depict slighly higher magnitudes, but still within the confidence bands.\footnote{This is in line with standard macro textbook \citep{Mankiw} and what central banks believe today. In the United States, the \href{https://www.federalreserve.gov/newsevents/pressreleases/monetary20221102a.htm}{November 2022 FOMC statement} indicates that the Committee will consider lags as it determines the pace of future increases in the federal funds rate. This is also true for other central banks. For instance, Philip R. Lane, a Member of the Executive Board of the ECB, in a recent speech at the Conference \href{https://www.ecb.europa.eu/press/key/date/2022/html/ecb.sp221011~5062b44330.en.html}{\textit{EU and US Perspectives: New Directions for Economic Policy}}, has commented on the lagged effect of monetary policy transmission to the real economy.}
All the other real variables follow a similar path, even though no restrictions on the shape of the impulse responses are imposed. This suggests that monetary policy shocks are transmitted to the real economy with a lag of a few months, in line with what monetary policymakers believe. Indeed, both real consumption, real income, and capacity utilization decline in a hump-shaped manner following the shock, reaching the maximum impact just before the end of the first year before returning to their trend, with a peak magnitude ranging between -1\% and -3\%. Real consumption, on the other hand, seems to react more quickly to the shock, which can be explained by the sharp drop in house prices, a mechanism well documented in the literature \citep[e.g.,][]{mian2013household,slacalek2020household}, a result we will comment on later. At the same time, business sales, business inventories, and new orders for durable goods all contract in line with industrial production. Turning to the labor market, the unemployment rate shows a sluggish increase, with no response at impact. The increase starts only from the second month onwards and it reaches its peak around 0.5\% after about ten months \citep[see e.g.,][]{christiano1999monetary}. The average real earnings do not seem to react significantly to a monetary shock at impact, but progressively decrease over time,  confirming the rather sluggish nature of real wages and possibly suggesting the presence of frictions in the economy, an idea that is now found in almost all the standard macro-theoretical models.

\subsubsection*{Housing Market}
The analysis of both housing market and financial variables can further help us in better understanding which other channels are at play in the propagation of the monetary policy shock. Housing investments reduce at a large magnitude following a monetary policy tightening, with both housing starts and new housing permits which shrink at impact by around 10\%, reaching almost -20\% at peak within six months, though they also exhibit a rather short-living response. Coherently, house prices decrease by 0.5\% at impact for GK, MAR and BS, and all the instruments point to a more long-lasting effect compared to the other housing sector indicators. Overall, the high sensitivity of the housing market to monetary policy shocks may directly impact household balance sheets, as also confirmed by the sharp reduction in consumption expenditure. As underlined by recent studies, this channel might be more relevant than previously believed. Indeed, when combining the large percentage of real estate over total assets in the portfolio of households with low level of wealth \citep{franconi2022} along with the large marginal propensity to consume (MPC) of \textit{hand-to-mouth} households \citep[e.g. see][]{kaplan2018microeconomic} and the larger MPC in response to a negative income shock \citep{christelis2019asymmetric}, the household balance sheets channel takes on a greater relevance for the transmission of monetary policy shocks \citep[see for instance][]{slacalek2020household}.

\subsubsection*{Financial Market}
The negative wealth effect coming from the housing market can potentially compoud with that coming from the financial market, further contributing to the amplification of the monetary policy shock. In the literature, these mechanisms are referred to as the financial accelerator and the credit channel \citep{bernanke1995inside,bernanke1999financial}. Asset prices, proxied by the S\&P 500, show a sudden and large repricing for the Gk isntrument, experiencing a decline of more than 10\% already in the second month. Interesting, the other instruments also show a significant drop, equal to 5\%, which is however half compared to what implied by GK. More generally, all financial variables included, such as the Moody's Corporate Bond Spread (BAA-AAA) and all interest rates across different maturities, react at impact. Interest rates rise across all maturities, albeit to different degrees, with the ten-year Treasury yield which rises less compared to the federal funds rate and the one-year Treasury yield\. This is reflected in an inversion of the yield curve and a negative reaction of the term spread, which reduces by around 50bp in the first months. Turning to the credit channel, the tightening of financial conditions drives an increase in reserves and the corresponding decline in credit throughout the economy. Business loans, real estate loans, and nonrevolving consumer credit fall sluggishly, reaching a trough of roughly 2 to 4 per cent the former and 1 to 3 percent the latter about 20 months after the shock.

\subsubsection*{Exchange Rates}
In terms of exchange rates, the US dollar appreciates following a monetary policy tightening, as shown by the GBP/USD, CAD/USD, CHF/USD and JPY/USD exchange rates, which increase by roughly 4\% (2\% for JPY) in the first two months. This implies higher prices for those countries that import goods produced in the US, and thus a decrease in US exports, which can further negatively weigh on the overall impact of the monetary policy shock on the domestic economy. It is also worth noting that both exchange rates react quickly, peaking within the second month before gradually declining. Thus, they do not exhibit the \textit{delayed overshooting puzzle} that was instead presented by the VAR analysis of \cite{eichenbaum1995some}, a result that confirms what was also found by \cite{forni2010dynamic}.

\subsubsection*{Prices}
Finally, a monetary policy tightening unequivocally reduces prices. All of the different price measures share a similar pattern: the CPI, the PCE deflator, the PPI, house prices and oil prices all react at impact for almost all the instruments, a feature that is at odds with the classical recursive identification scheme, which assumes zero contemporaneous effects at impact. Although the contraction is immediate, prices do not fully adjust at impact, but continue to fall for several months. The large fall in oil prices observed with GK, MAR and BS instruments could potentially be explained by a fall in consumption and investment following the negative monetary policy shock, which then affects oil prices through the demand channel by directly reducing the demand for oil.

\subsubsection*{Global Spillovers}
A US monetary policy shock plays also as an important role in terms of global spillovers, thus contributing in shaping the global outlook. As shown by many studies, for instance \cite{ca2020monetary}, many of the channels which we analysed in the above paragraph, such as the demand channel, or the exchange channel, contribute to the propagatation of negative spillovers to other countries' financial markets and real activity sectors. Though representing an interesting topic, our analysis is not strictly related to the global dimension of US monetary policy, which may be explored in future works.

\begin{center}
	\begin{table}[h!]
	\centering
		\caption{Forecast Error Variance Decomposition: Proxy DFM (GK)}
		\medskip
		\renewcommand{\arraystretch}{.9}
\begin{tabular}{lccccccc}
\hline\hline
\textbf{Variables} & & \textit{h=0} & \textit{h=6}  & \textit{h=12}  & \textit{h=18}   & \textit{h=24} & \textit{h=30} \\
\hline
 Industrial production && 0.37 & 10.46 & 22.07 & 25.02 & {\color{dkred}26.18} & 26.10  \\
Real Consumption && 9.07 & 30.65 & {\color{dkred}32.65} & 31.53 & 30.20 & 29.23  \\
Real Income ex.Trans. && 7.01 & 10.32 & 21.42 & {\color{dkred}23.41} & 22.74 & 21.03  \\
Business Sales && 3.68 & 29.11 & {\color{dkred}35.67} & 35.61 & 34.67 & 33.54  \\
New Orders: Durables && 4.76 & 27.16 & 33.35 & 34.41 & {\color{dkred}34.71} & 34.27  \\
Business Inventories && 8.72 & 0.95 & 5.89 & 10.38 & 13.34 & {\color{dkred}14.62}  \\
Cap. Util. Manuf. && 0.04 & 13.03 & 24.49 & 26.80 & {\color{dkred}27.24} & 26.56  \\
Business Sentiment && 1.37 & 37.49 & {\color{dkred}38.57} & 31.22 & 26.39 & 24.32  \\
Unemployment rate && 2.53 & 16.61 & 26.65 & 29.82 & {\color{dkred}30.33} & 28.84  \\
Avg. Hours manuf. && 1.21 & 17.70 & {\color{dkred}27.35} & 27.13 & 26.35 & 26.70  \\
Avg. Earnings manuf. && {\color{dkred}23.54} & 5.44 & 9.79 & 9.68 & 9.46 & 9.24  \\
CPI Headline && {\color{dkred}15.28} & 7.92 & 4.30 & 3.06 & 2.97 & 3.42  \\
PCE && {\color{dkred}14.74} & 9.19 & 5.26 & 3.79 & 3.60 & 3.96  \\
PPI && {\color{dkred}18.04} & 13.23 & 8.21 & 6.24 & 5.63 & 5.60  \\
Oil Price && {\color{dkred}22.58} & 18.25 & 12.29 & 9.75 & 8.66 & 8.20  \\
House Price && 20.15 & {\color{dkred}22.16} & 19.09 & 18.04 & 17.26 & 15.99  \\
Housing Starts && 30.40 & {\color{dkred}61.10} & 46.70 & 33.01 & 25.14 & 21.88  \\
New Housing Permits && 29.35 & {\color{dkred}60.44} & 45.46 & 32.62 & 26.14 & 24.12  \\
Total Reserves && {\color{dkred}45.03} & 38.35 & 31.00 & 27.80 & 26.27 & 25.41  \\
Business Loans && {\color{dkred}72.01} & 17.90 & 12.38 & 16.66 & 19.47 & 20.08  \\
Real Estate Loans && {\color{dkred}53.88} & 10.97 & 15.51 & 19.63 & 21.29 & 20.99  \\
One-year rate && {\color{dkred}39.21} & 18.39 & 14.20 & 12.18 & 11.68 & 11.50  \\
Ten-year rate && {\color{dkred}28.27} & 11.64 & 8.81 & 7.27 & 6.38 & 5.89  \\
BAA-AAA && {\color{dkred}35.12} & 16.17 & 12.41 & 10.14 & 8.99 & 8.38  \\
S\&P 500 && 51.54 & {\color{dkred}53.53} & 40.85 & 34.98 & 31.86 & 29.93  \\
Exch. GBP/USD && {\color{dkred}71.65} & 67.14 & 60.74 & 59.67 & 58.08 & 54.28  \\
Exch. CAD/USD && 60.88 & {\color{dkred}61.73} & 51.05 & 46.80 & 44.83 & 43.93  \\
Exch. CHF/USD && 54.56 & {\color{dkred}58.47} & 53.40 & 47.51 & 41.80 & 36.11  \\
Exch. JPY/USD && 17.84 & 22.43 & {\color{dkred}23.14} & 21.19 & 18.74 & 16.32  \\
\hline\hline
\end{tabular}

		\subcaption*{Notes: Forecast Error Variance Decomposition of the contractionary monetary policy shock for a selection of varibales, at different horizons. The shock in the Proxy DFM is identified with the instrument GK. Values in red represents the peak relevance.}
		\label{MP_VD}
	\end{table}
\end{center}

\subsubsection*{Variance Decomposition}
Table \ref{MP_VD} presents the variance decomposition analysis obtained from the GK instrument and confirms that a monetary policy shock has an important role in explaining the cyclical fluctuations of the economy. For real variables and indicators of economic activity, the shock explains a small fraction of the variance at impact, ranging from 1\% for industrial production to 9\% for real income. This is consistent with the impulse response analysis, where a monetary policy shock has a limited effect on most of the real variables at the very impact, with the shock taking a few months to be transmit to the economy. However, its importance increases at longer frequencies, generally peaking during the course of the second year at roughly 30\%. Similar results are shared by labor market variables, specifically for the unemployment rate and hours worked. The variance explained of the average earnings seems, however, to remain rather low, suggesting what already observed in the IRFs. Turning to the nominal variables, this analysis shows a larger share of variance explained at the beginning of the forecast horizon, from 14.7\% of PCE to 22.5\% of oil prices, with the importance of the monetary policy shock dissipating at longer horizons. On the housing and financial market, conversely, a monetary shock explains a large fraction of the total variance of these variables within the first six months following the shock. The peak in variance explained by the monetary policy shock for the housing sector is around 60\%, for the credit market in the range 53\% to 72\%, for interest rates between 28\% to 35\%, for S\&P 500 index above 50\% and quite persistent, for exchange rates in the range 54\% to 71\%, with the exception of the $JPY/USD$ which amounts to almost 18\%.\footnote{The Bank of Japan conducts interventions in the foreign exchange rate market under the input of the Ministry of Finance. This may explain why US monetary policy shocks are comparatively less relevant for changes in the JPY/USD exchange rate.}

\section{Conclusion}
\label{sec:conclusion}

External instruments identification procedure is not safe from issues affecting traditional SVARs. Even if the instrument is perfect, the estimates can be biased.

By means of a theoretical model with perfect foresight, we show that, if the underlying shock is non-fundamental or the variables are observed with a measurement error, the SVAR consistently fails to estimate the true impulse responses. Moreover, subjective choices about the variables included in the model can further increase the uncertainty in both the magnitude and the sign of the estimated responses. The latter is a problem that also affects FAVARs. As a solution, we propose using external instruments in a DFM, which is able to address all mentioned issues at once and to estimate the correct IRFs.

In the empirical exercise, we focus on an application to monetary policy and consider  the most well-known monetary policy instruments in the literature. Results show that, unlike SVAR, the information included in the DFM is enough to estimate puzzle-free responses in line with economic theory. Interestingly, results and consistent regardless the instrument considered, suggesting that the larger information set is able to deal with the distorting effect of monetary policy news.

Moreover, DFM proves invaluable in examining the behavior of a large set of variables, simultaneously. A tool of great value, especially for central banks, which allows for an internally consistent examination of the transmission channels of monetary policy. Our analysis shows that a monetary policy tightening shock has a clear contractionary effect on the economy, leading to a decline in both economic activity and prices. Multiple channels come into play, with both the financial and housing sectors deteriorating and directly affecting private consumption. Finally, the variance decomposition analysis shows that the monetary policy shock explains a significant portion of the variance of both real and nominal variables, albeit at different horizons, further highlighting the role of monetary policy in influencing business cycle fluctuations.

\clearpage

\setlength{\bibsep}{0pt plus 0.5ex}
\bibliographystyle{aer}
{\small \bibliography{biblio}}

\clearpage

\begin{appendices}
\section{Absolute shock and IRFs}
\label{sec:unitvarshock}

If we have fundamentalness, the reduced-form residuals, say $u_t$, can be written as a linear combination of the structural shocks, say $\eta_t$. Formally:
\begin{align}
	u_t= & Q\eta_t\\
	\intertext{The external instrument identification allows to obtain covariance restrictions from proxies for the latent structural shock of interest in line with conditions (\ref{condA}) and (\ref{condB}). We proceed as usual with the first-stage regression $z_t=\delta^{'}u_t+\nu_t$. Then, we have that for the shock $i$:}
	\delta = & \frac{\mathbb{E}[z_t u_t]}{\Sigma_u} = q_i \frac{\mathbb{E}[z_t\eta_{it}]}{\Sigma_u} = q_i\frac{\alpha}{\Sigma_u}\\
	\intertext{where $q_i$ is the $i$-th column of the matrix $Q$, which corresponds to the structural shock of interest $i$ and is then equal to $q_i=\frac{\Sigma_u}{\alpha}\delta$, and $\Sigma_u$ is the variance-covariance matrix of the reduced-form innovations (or common shocks) which is equal to the identity matrix in a DFM framework. \cite{forni2022external} show that if the shock is fundamental then we can estimate consistently $\alpha$ as follows:}
	\hat{\alpha} = & \sqrt{\hat{\delta}^{'}\hat{\Sigma}_u\hat{\delta}}		
	\intertext{Hence, the absolute (unit-variance) structural shock $i$ is:}
	\eta_{it}=& q_i^{'}\Sigma_u^{-1}u_t =  \frac{\delta u_t}{\sqrt{\delta^{'}\Sigma_u \delta}}\\
	\intertext{and the corresponding impulse response functions to the structural shock $i$ are:}
	c_i(L) = &B_{\chi}(L)\Sigma_u q_i = B_{\chi}(L)\Sigma_u \frac{\delta}{\sqrt{\delta^{'}\Sigma_u \delta}}
\end{align}

\section{On the Instrument}
\label{sec:dirtyinsrtu}

In the previous sections we adopted a rather strong condition and we assumed the instrument available to the econometrician to be \textit{perfect}, \textit{i.e.} to be the true structural shock itself. However, as originally pointed out by \cite{mertens2013dynamic}, the instrument is very likely to be measured with an error, which, moreover, could also compound with other sources of distortion. Therefore, in what follows we relax some of the conditions assumed so far, studying how different models behave in a more generic framework.

Following \cite{plagborg2022instrumental} and \cite{forni2022external}, we generalize the data generating process of the instrument, which is now represented by the equation (see also the discussion in \cite{stock2018identification}):

\begin{equation}
	z_t = \phi(L)z_{t-1} + \lambda(L) y_{t-1} + \alpha\varepsilon_{1,t} + \epsilon_t \label{instr}
\end{equation}

where $\epsilon_t \sim N(0,\sigma_{\epsilon})$, $\alpha \geq 0$ is a scalar measuring (along with  $\sigma_{\epsilon}$) the overall strength of $z_t$ as IV and $y_{t-1}$ are past observations of the endogenous variables. $\phi(L)$ and $\lambda(L)$ are rational polynomials in the lag operator, $L$. 

One can immediately notice how using the shock of interest itself as a direct instrument represent a special case of equation (\ref{instr}). In other words, in the previous sections we were implicitly assuming the following assumptions to be satisfied:
\begin{enumerate}[label=\textit{A.{\arabic*}}]	 	
	\item \textit{The IV does not depend on past values, i.e. $\phi(L) = \lambda(L) = 0 $} \label{a1}
	\item \textit{The IV is perfectly related to the shock of interest, i.e. $\alpha = 1$} \label{a2} 
	\item \textit{The IV is measured without error, i.e. $\sigma_{\epsilon}=0$}. \label{a3}
\end{enumerate}

Let us briefly discuss their implications. Assumption \ref{a1} is what \cite{stock2018identification} define as ''lag exogeneity assumption''. Although rather stringent in an LP-IV setting, it is not a strong requirement for the SVAR-IV methodology. Nevertheless, we will also consider the case in which this assumption does not hold.  To understand the implications of \ref{a2}-\ref{a3}, we instead focus on the signal-to-noise ratio. Without loss fo generality, let us consider a situation in which only \ref{a1} holds, so that the ratio is $\alpha^2/\sigma_{\epsilon}^2$, and the value of the constant $\alpha$ measures the correlation between the instrument and the structural shock. Clearly, when $\alpha = 0$ the ratio is zero and, thus, the instrument is of small interest because it is not correlated with the observable variable. On the other hand, the higher $\alpha$, the greater the importance of the shock (ceteris paribus). Analogously, as $\sigma_{\epsilon}$ tends to zero the information that the instrument conveys is perfect up to the scale $\alpha$, as the ratio goes to infinity. 

To better understanding the importance of \ref{a1}-\ref{a2}-\ref{a3} and their implications in the estimation of structural shocks, we perform a simulation exercises in which we compare estimates obtained with four instruments of different \textit{quality}. The four different instruments  are obtained from equation \ref{instr} by switching on-and-off some of the assumptions described in the previous paragraph, specifically:

\begin{align*} 
	I1:& z_t = \tilde\alpha^2\varepsilon_{1,t} \\
	I2:& z_t = \tilde\alpha^2\varepsilon_{1,t} + \epsilon_t \\
	I3:& z_t = \tilde\alpha^2\varepsilon_{1,t} + \phi z_{t-1}         \\	
	I4:& z_t = \tilde\alpha^2\varepsilon_{1,t} + \tilde\delta' y_{t-1}  
\end{align*}

with $\tilde \alpha \sim N(0,1)$ and $\sigma_{\epsilon} \sim U(0,0.5)$, whereas the other parameters are deterministic and arbitrarily calibrated following \cite{forni2022external} as $\phi = 0.5$ and $\tilde \delta = [-0.6 \quad 0.4 \quad 0]'$. 

\begin{figure}[h!]
	\caption{Comparison of weak instruments}
	\subcaptionbox{VAR}{
		\includegraphics[trim=4.55cm 0cm 3.6cm 0cm,clip=true,width=1\textwidth]{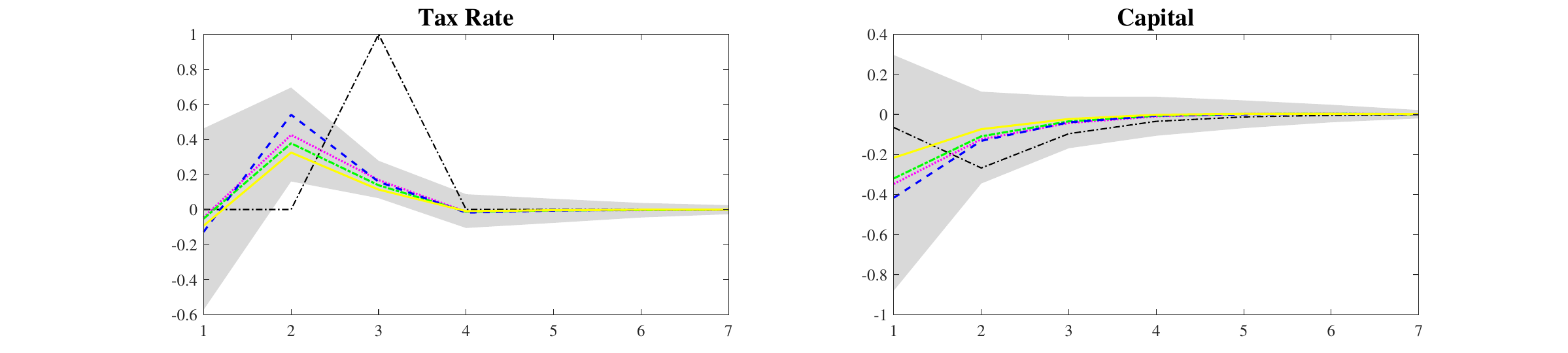}}
	\subcaptionbox{DFM}{
		\includegraphics[trim=4.55cm 0cm 3.6cm 0cm,clip=true,width=1\textwidth]{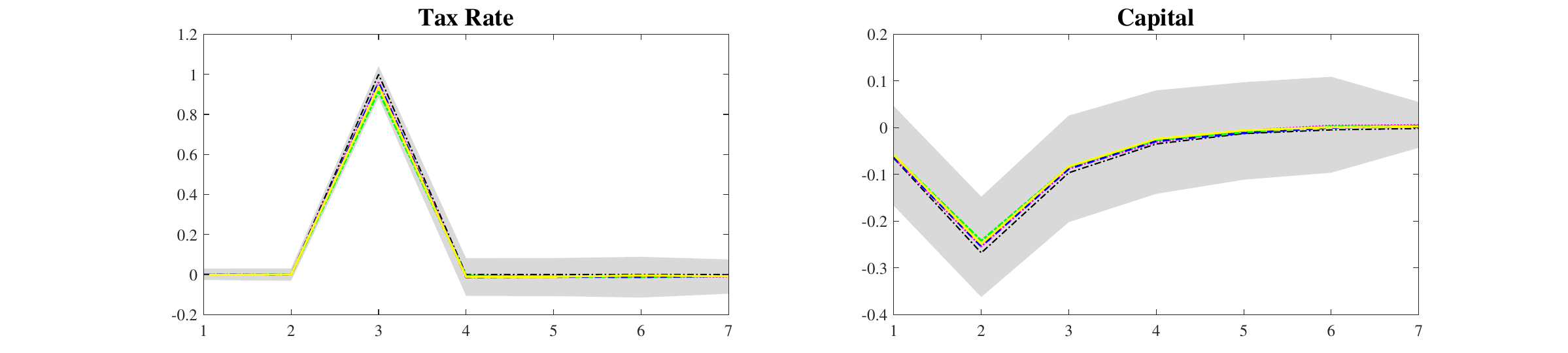}}
	\medskip
	\subcaption*{Notes: the two panels show the responses of tax rate and capital to a tax shock. Panel (a) is obtained with a fundamental trivariate VAR with tax reate, capitla and technology but augemnted with a small measurement error. Panel (b) is obtained with the Proxy DFM on the entire dataset generated with a measuremnt error. In both cases, the measurement is generated with $\nu = 0.5$. In each panel, the black dashed lines are the true response, in magenta the reposnes for $I1$, in green the responses for $I2$, in blue the responses for $I3$ and in yellow for $I4$. }
	\label{weaks}
\end{figure}

The instruments are then used in estimating the underlying structural shocks in two models, which consist in a trivariate (and fundamental) VAR augmented with measurement error and a DFM on the entire dataset generated with a measuremnt error. For the construction of both the models we follow the procedure described in the previous section. Results are reported in Figure \ref{weaks}, and confirm what already found in previously. Indeed, the overall picture does not change with respect to the perfect instrument case and results are robust also over different specification of the instrument genereting process. Not surprisinly, the trivariate VAR (Panel a) systematicallty fails to estimate the structural shock in neither the cases even if fundamental, with the overall results suffering from the presence of the measurement error. Conversely, the DFM (Panel b) is able to recover the true responses almost perfectly: results do not exhibit any bias and are robust across all the four cases studied, thus further confirming the goodness of this model and its good match with the proxy identification technique.

\clearpage

\section{Tables}

\vfill

\begin{center}

	\begin{table}[h!]
		
		\centering
		\caption{Frobenius Norm between theoretical and empirical IRF [Shocks]}
		\medskip
		\begin{tabular}{llcccc}
\hline\hline
& & \multicolumn{2}{c}{Fiscal Foresight} & \multicolumn{2}{c}{No Fiscal Foresight}\\ \cline{3-6}
$\nu$ & Model &\textbf{Shock}&\textbf{IRF}&\textbf{Shock}&\textbf{IRF}\\\hline
$0.5$& \textbf{VAR}&100.00&100.00&100.00&100.00\\
& \textbf{FAVAR}&20.72&16.56&99.34&130.72\\
& \textbf{DFM}&17.95&15.07&102.48&130.61\\\cline{3-6}
$2$& \textbf{VAR}&100.00&100.00&100.00&100.00\\
& \textbf{FAVAR}&31.19&20.94&60.10&77.43\\
& \textbf{DFM}&19.86&16.04&57.08&73.40\\\cline{3-6}
$5$& \textbf{VAR}&100.00&100.00&100.00&100.00\\
& \textbf{FAVAR}&39.85&23.47&53.99&53.66\\
& \textbf{DFM}&23.90&17.94&44.52&47.71\\
\hline
\hline
\end{tabular}

		\subcaption*{}
		\label{frob_full}
	\end{table}
\end{center}

\begin{center}
	\begin{table}[h!]
		
		\centering
		\caption{Fundamentalness Test}
		\medskip
		\begin{tabular}{lcccccc}
\hline\hline
& & \multicolumn{5}{c}{Leads}\\
\cline{3-7}
Model& $\alpha$ &\textbf{1}&\textbf{2}&\textbf{3}&\textbf{4}&\textbf{5}\\\hline
\textbf{VAR}   & 1\%&64.29&85.71&85.71&89.80&88.78\\
&    5\%&75.51&87.76&88.78&89.80&89.80\\
&    10\%&76.53&89.80&91.84&91.84&91.84\\
\cline{3-7}
\textbf{FAVAR} & 1\%&64.29&59.18&54.08&53.06&52.04\\
&  5\%&67.35&63.27&57.14&55.10&54.08\\
&  10\%&68.37&64.29&60.20&57.14&56.12\\
\cline{3-7}
\textbf{DFM}   & 1\%&0.00&0.00&0.00&0.00&0.00\\
&    5\%&0.00&0.00&0.00&0.00&0.00\\
&    10\%&0.00&0.00&0.00&0.00&0.00\\
\hline
\hline
\end{tabular}

		\label{fund_simu}
	\end{table}
\end{center}

\clearpage

\section{Figures}

\vfill

\begin{figure}[h!]
	\centering
	\caption{Comparison models specifications}
	\includegraphics[trim=3.8cm 0cm 3.5cm 0cm,clip=true,width=1\textwidth]{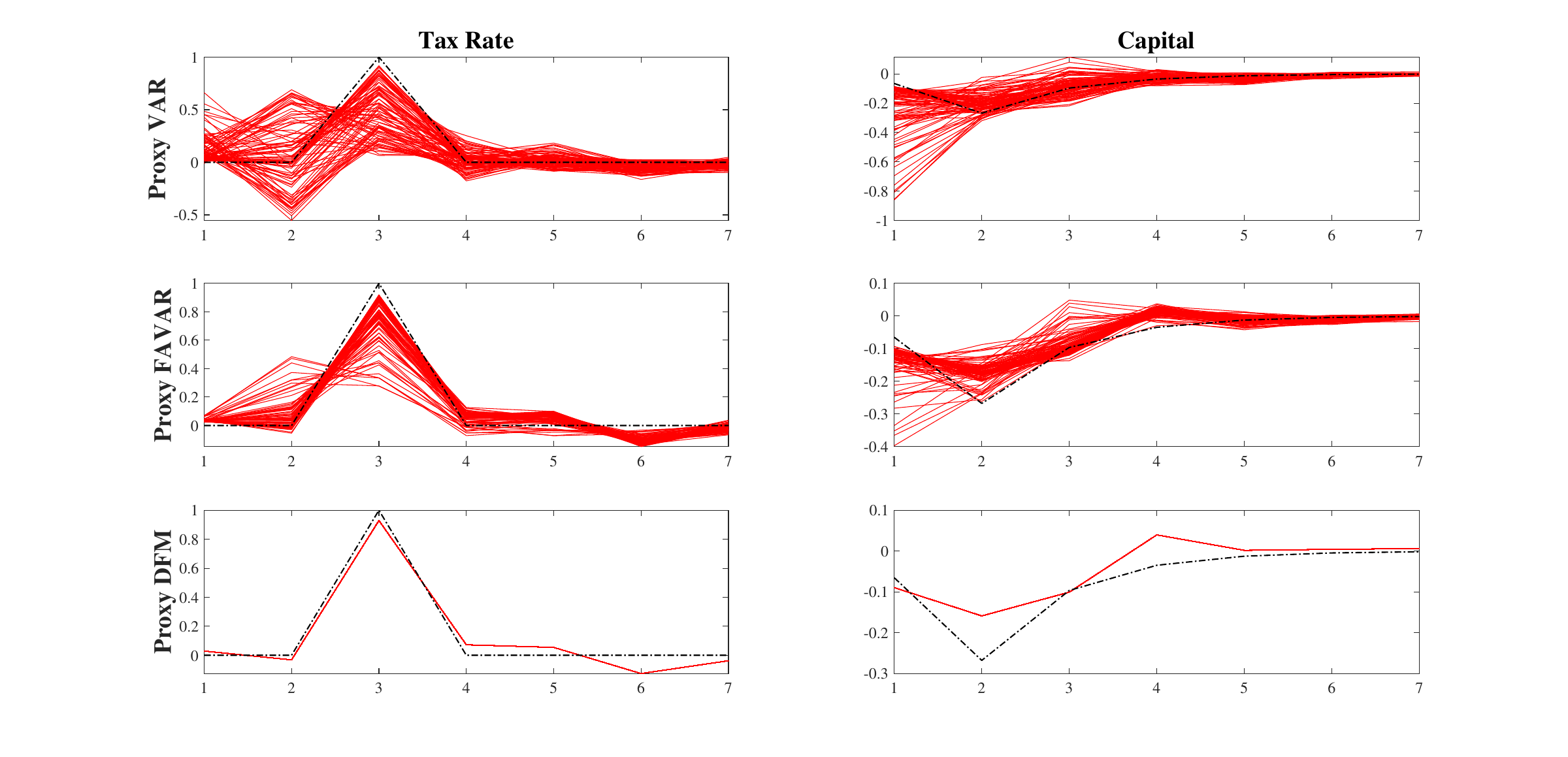}
	
	\subcaption*{Notes: black dotted lined are the true responses, the red lines the empirical responses obtained with different specificatio of the models. The resposnes are computed as point estimates of the IRFs obtained with different specification of the model. Going from the upper panels to the lower, we see results obtained with VAR(2) on capital, tax rate and a third variable, with FAVAR(2) on capital, tax rate, a third variable and two factors and with a DFM(2). The third variable for VAR(2) and FAVAR(2) varies at each iteration. DFM(2) include all the variables in the estimation.}
	\label{modelspec}
\end{figure}

\vfill


\clearpage


\vfill

\begin{figure}
	\centering
	\caption{Comparison of the shocks across models and $\nu$}
	
	\includegraphics[trim=4cm 0cm 3.5cm 0cm,clip=true,width=1\textwidth]{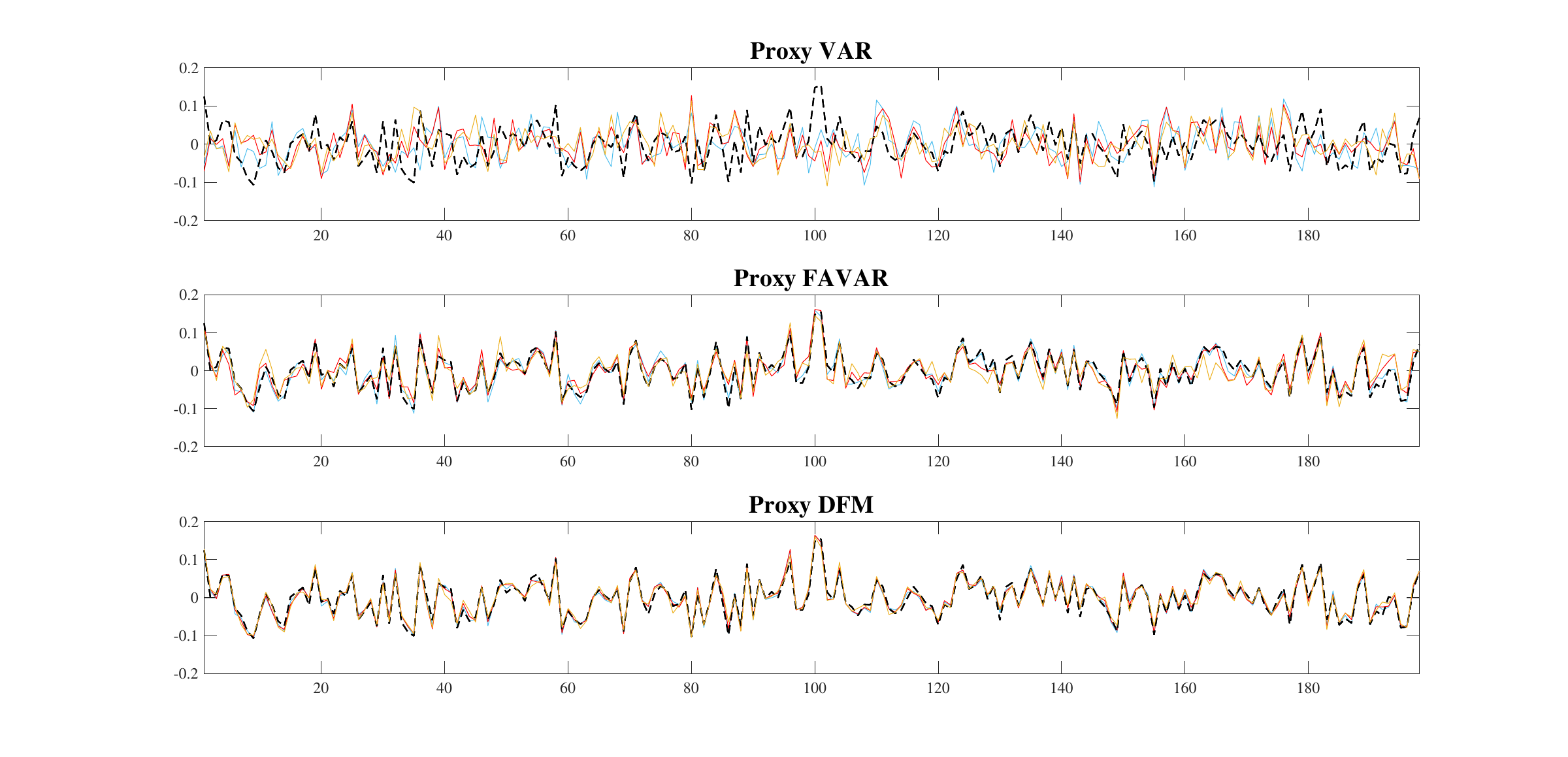}
	\subcaption*{Notes: Comparison between the mean of unit-variance shocks across models and different values of $\nu$. Black dotted lines are the "true" responses. Responses obtained in case $\nu = 0.5$ are displayed in blue, along with their $68\%$ confidence bands in grey. The red lines are the response obtained in case $\nu = 2$. The yellow lines are the response obtained in case $\nu = 5$. The series are computed as sample average of the shocks obtained across simulations. 
	}

	\label{shocks_by_models}
\end{figure}

\vfill
\clearpage

\vfill
\begin{center}

	\begin{figure}[h!]
		
		\centering
		\caption{IRF of tax rate and capital to a tax shock, with no fiscal foresight.}
		\includegraphics[trim=3.8cm 0cm 3.5cm 0cm,clip=true,width=1\textwidth]{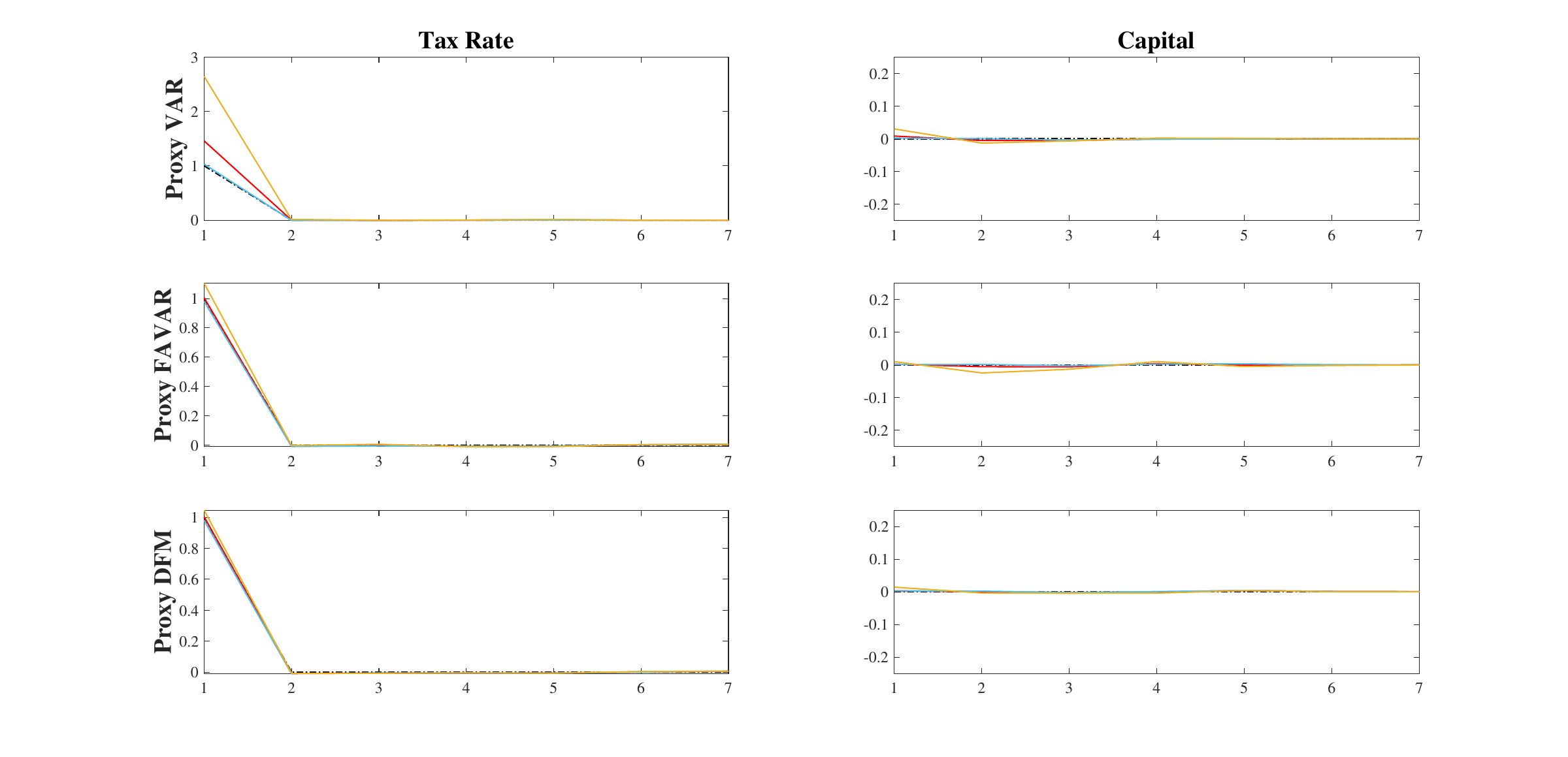} 
		\medskip
		
		\subcaption*{Note: Black dotted lines are the "true" responses. Responses obtained in case $\nu = 0.5$ are displayed in blue, along with the $68\%$ confidence bands in grey. The red lines are the response obtained in case $\nu = 2$. The yellow lines are the response obtained in case $\nu = 5$. In each case the external intrument is the real structural shock $u_{\tau,t}$.}
		\label{simuclean_NFF}
		
	\end{figure}
\end{center}
\vfill

\clearpage

\begin{figure}[h!]
	\centering
	\caption{VAR vs DFM with GK identification: shock comparisonn}
	\includegraphics[trim=4cm 0.5cm 3.5cm 0cm,clip=true,width=1\textwidth]{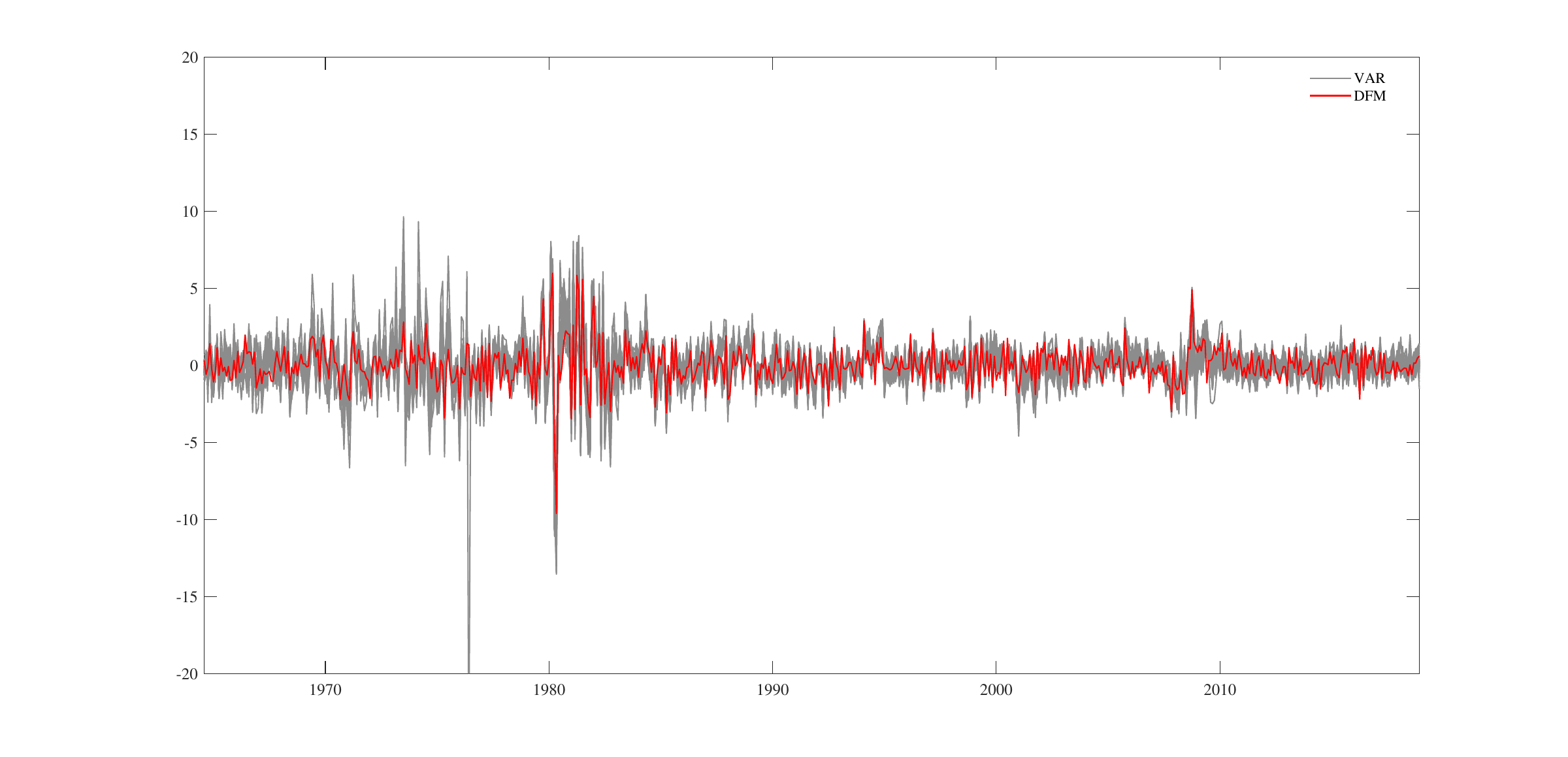}
	
	\subcaption*{Notes: Comparison between the structural monetary policy shocks estimated from the 95 different VAR specifications (grey lines) and the DFM (red line). The structural shock has been recovered via external instrument taken by Gertler and Karadi.}
	\label{VAR-DFM_shock}
\end{figure}

\clearpage

\clearpage

\section{Robustness}
\label{sec:robustness}


\begin{figure}[h!]
	\caption{Robustness on the number lags $p$} 
	
	\includegraphics[trim=4cm 0.5cm 3.5cm 0cm,clip=true,width=1\textwidth]{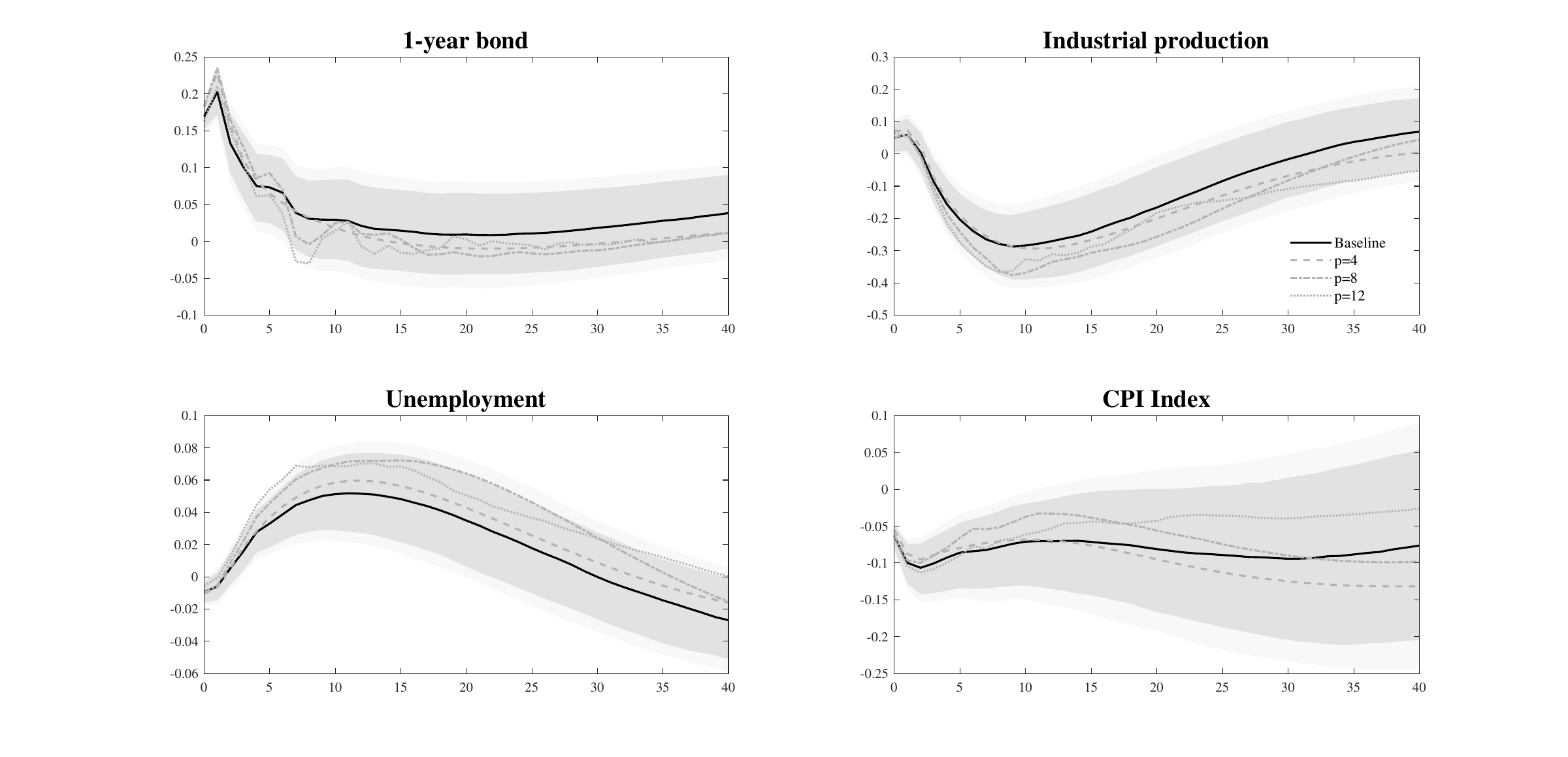}
	
	\subcaption*{Notes: Comparison of responses to the shock identified using GK instrument with different lag lengths $p$. Dark and light grey shaded areas are respectively the $84\%$ and $90\%$ confidence bands for the baseline (i.e. $p = 8$). Bands are obtained with bootstrap techniques.}
	\label{rob_p}
\end{figure}

\begin{figure}[h!]
	\caption{Robustness on the number of static factors $r$}
	
	\includegraphics[trim=4cm 0.5cm 3.5cm 0cm,clip=true,width=1\textwidth]{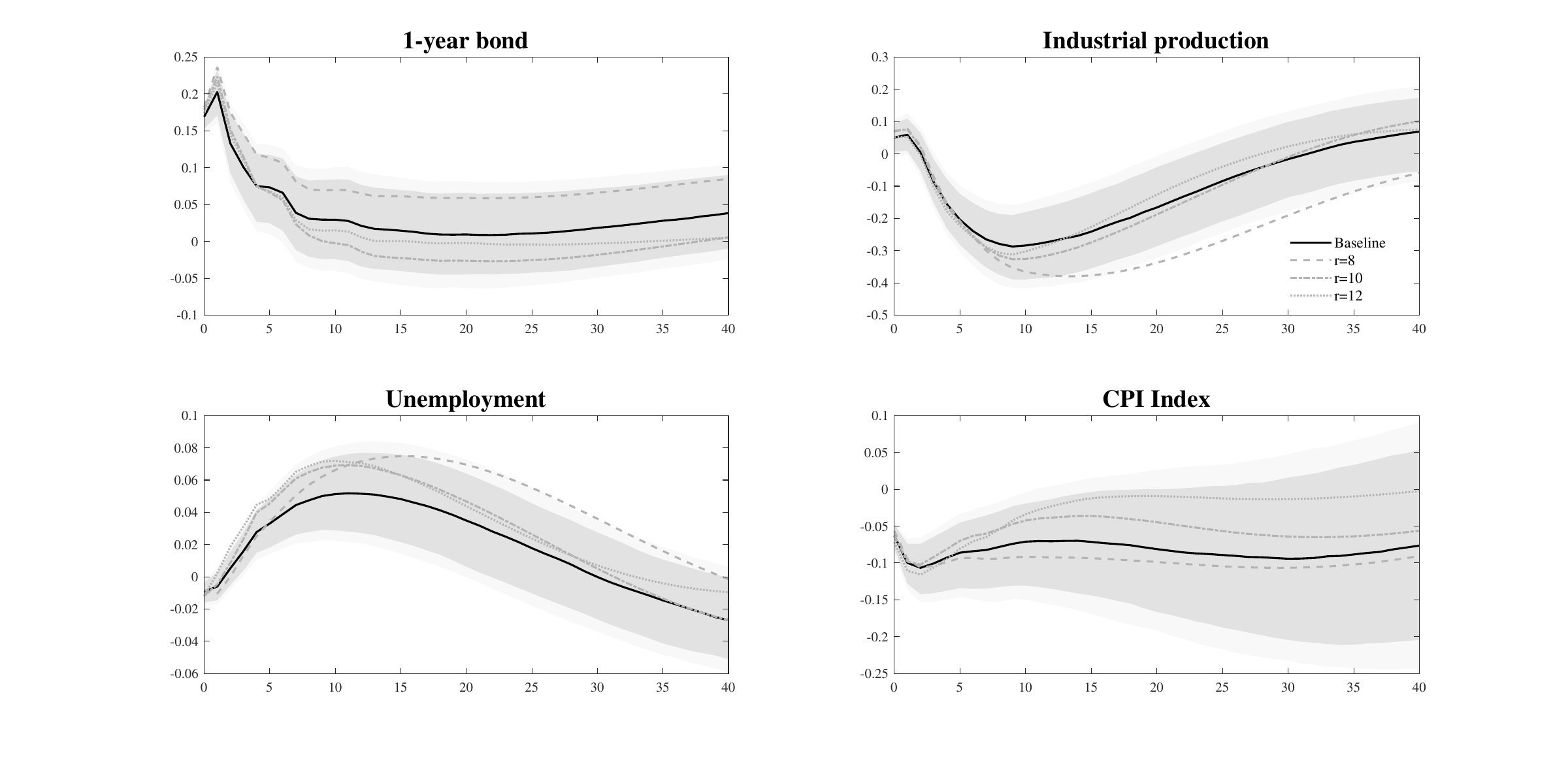}

	\subcaption*{Notes: Comparison of responses to the shock identified using GK instrument with different number of static factors $r$. Dark and light grey shaded areas are respectively the $84\%$ and $90\%$ confidence bands for the baseline (i.e. $r = 9$). Bands are obtained with bootstrap techniques.}
	\label{rob_r}
\end{figure}

\begin{figure}[h!]
	\caption{Robustness on the number of dynamic factors $q$}
	\includegraphics[trim=4cm 0.5cm 3.5cm 0cm,clip=true,width=1\textwidth]{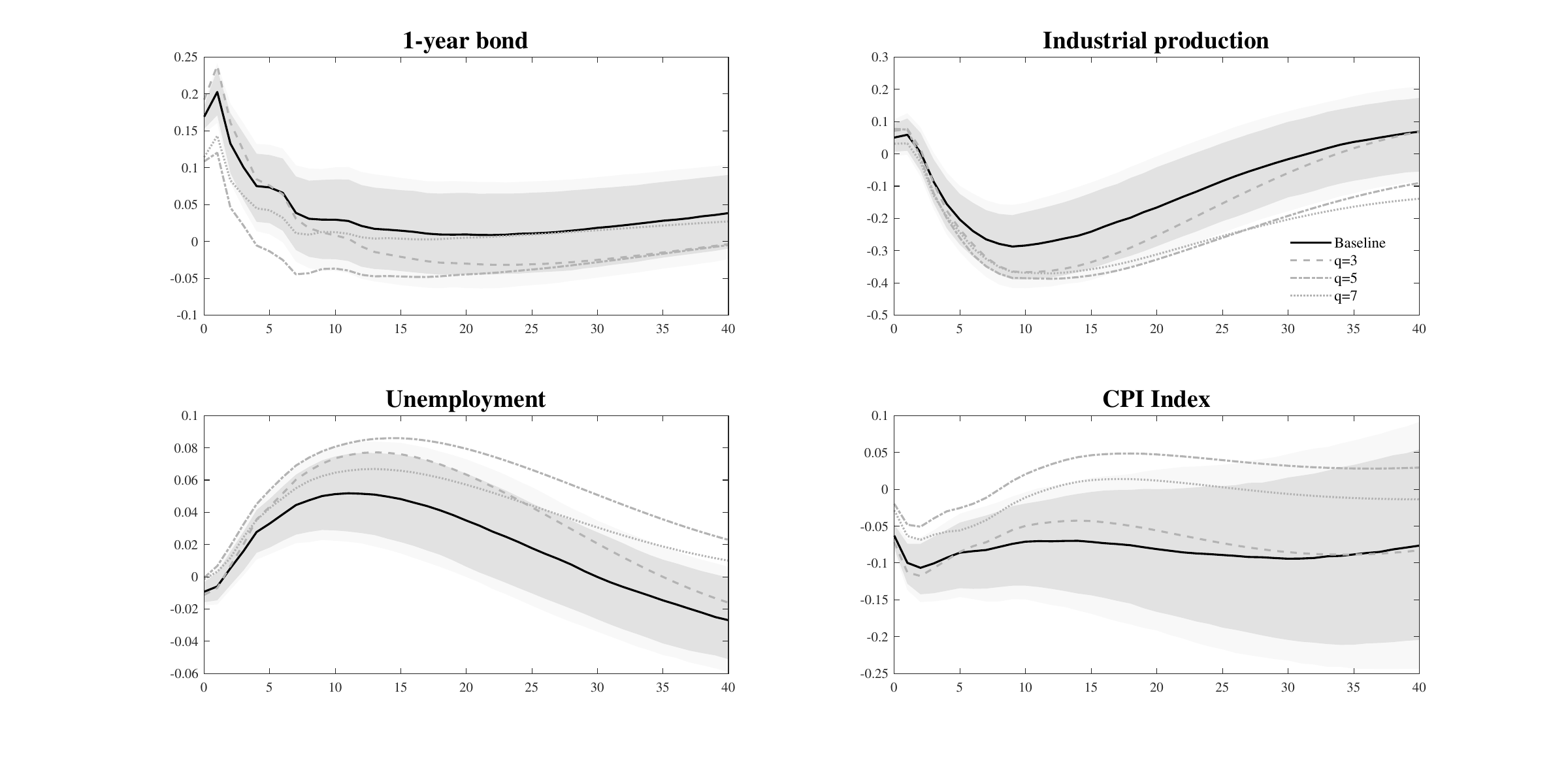}
	\subcaption*{Notes: Comparison of responses to the shock identified using GK instrument with different number of dynamic factors $q$. Dark and light grey shaded areas are respectively the $84\%$ and $90\%$ confidence bands for the baseline (i.e. $q = 4$). Bands are obtained with bootstrap techniques.}
	\label{rob_q}
\end{figure}
\end{appendices}

\end{document}